\documentclass[twocolumn, trackchanges]{aastex631}
\usepackage{multirow}
\usepackage{booktabs}
\usepackage{ulem}
\usepackage{url}
\usepackage{color}
\newcommand{\f}{\phantom{2}}

\newcommand{\ltsimeq}{\raisebox{-0.6ex}{$\,\stackrel 
        {\raisebox{-.2ex}{$\textstyle <$}}{\sim}\,$}} 
\newcommand{\gtsimeq}{\raisebox{-0.6ex}{$\,\stackrel 
        {\raisebox{-.2ex}{$\textstyle >$}}{\sim}\,$}}

\newcommand{\neiii}{[Ne\,{\sc iii}]}

\newcommand{\lya}{Ly$\alpha$}

\newcommand{\hbeta}{H$\beta$}
\newcommand{\halpha}{H$\alpha$}
\newcommand{\oiii}{[O\,{\sc iii}]}
\newcommand{\oii}{[O\,{\sc ii}]}

\newcommand{\asec}{^{\prime\prime}}
\newcommand{\amin}{^{\prime}}

\newcommand{\myemail}{chris.willott@nrc.ca}

\def\hst{{\it Hubble Space Telescope~}}
\def\jwst{{\it James Webb Space Telescope~}}

\received{2021 November 16}
\revised{2022 January 17}
\accepted{2022 January 31}
\submitjournal{PASP}

\shorttitle{NIRISS II: Wide Field Slitless Spectroscopy}
\shortauthors{Willott et al.}

\begin{document}

\title{The Near Infrared Imager and Slitless Spectrograph for the  {\it James Webb Space Telescope} - II. Wide Field Slitless Spectroscopy}

\correspondingauthor{Chris J. Willott}
\email{\myemail}

\author[0000-0002-4201-7367]{Chris J. Willott}
\affil{NRC Herzberg, 5071 West Saanich Rd, Victoria, BC V9E 2E7, Canada}


\author[0000-0001-5485-4675]{Ren\'e Doyon}
\affil{D\'epartement de Physique and Observatoire du Mont-M\'egantic, Universit\'e de Montr\'eal, C.P. 6128, Succ. Centre-ville, Montr\'eal, H3C 3J7, Qu\'ebec, Canada}
\affil{Institut de Recherche sur les exoplan\`etes, Universit\'e de Montr\'eal, Qu\'ebec, Canada}

\author[0000-0003-0475-9375]{Loic Albert}
\affil{D\'epartement de Physique and Observatoire du Mont-M\'egantic, Universit\'e de Montr\'eal, C.P. 6128, Succ. Centre-ville, Montr\'eal, H3C 3J7, Qu\'ebec, Canada}
\affil{Institut de Recherche sur les exoplan\`etes, Universit\'e de Montr\'eal, Qu\'ebec, Canada}

\author[0000-0003-2680-005X]{Gabriel B. Brammer}
\affil{Cosmic Dawn Center, Niels Bohr Institute, University of Copenhagen, Juliane Maries Vej 30, DK-2100 Copenhagen, Denmark}

\author[0000-0001-9184-4716]{William V. Dixon}
\affil{Space Telescope Science Institute, 3700 San Martin Drive, Baltimore, MD 21218, USA}

\author[0000-0002-7989-2595]{Koraljka Muzic}
\affil{CENTRA, Faculdade de Ci\^{e}ncias, Universidade de Lisboa, Ed. C8, Campo Grande, P-1749-016 Lisboa, Portugal}

\author[0000-0002-5269-6527]{Swara Ravindranath}
\affil{Space Telescope Science Institute, 3700 San Martin Drive, Baltimore, MD 21218, USA}

\author[0000-0001-8993-5053]{Aleks Scholz}
\affil{SUPA, School of Physics \& Astronomy, University of St Andrews, North Haugh, St Andrews, KY16 9SS, United Kingdom}

\author[0000-0002-4542-921X]{Roberto Abraham}
\affiliation{Department of Astronomy \& Astrophysics, University of Toronto, 50 St. George Street, Toronto, ON M5S 3H4, Canada}
\affiliation{Dunlap Institute for Astronomy and Astrophysics, University of Toronto, 50 St George Street, Toronto, ON M5S 3H4, Canada}

\author[0000-0003-3506-5667]{\'Etienne Artigau}
\affil{D\'epartement de Physique and Observatoire du Mont-M\'egantic, Universit\'e de Montr\'eal, C.P. 6128, Succ. Centre-ville, Montr\'eal, H3C 3J7, Qu\'ebec, Canada}
\affil{Institut de Recherche sur les exoplan\`etes, Universit\'e de Montr\'eal, Qu\'ebec, Canada}

\author[0000-0001-5984-0395]{{Maru\v{s}a Brada\v{c}}}
\affiliation{Department of Physics and Astronomy, University of California, Davis, One Shields Ave, Davis, CA 95616, USA}

\author[0000-0002-5728-1427]{Paul Goudfrooij}
\affil{Space Telescope Science Institute, 3700 San Martin Drive, Baltimore, MD 21218, USA}

\author{John B. Hutchings}
\affil{NRC Herzberg, 5071 West Saanich Rd, Victoria, BC V9E 2E7, Canada}

\author[0000-0001-9298-3523]{Kartheik G. Iyer}
\affiliation{Dunlap Institute for Astronomy and Astrophysics, University of Toronto, 50 St George Street, Toronto, ON M5S 3H4, Canada}

\author[0000-0001-5349-6853]{Ray Jayawardhana}
\affiliation{Department of Astronomy, Cornell University, Ithaca, New York 14853, USA}

\author[0000-0002-5907-3330]{Stephanie LaMassa}
\affil{Space Telescope Science Institute, 3700 San Martin Drive, Baltimore, MD 21218, USA}

\author[0000-0003-3243-9969]{Nicholas Martis}
\affil{NRC Herzberg, 5071 West Saanich Rd, Victoria, BC V9E 2E7, Canada}
\affil{Institute for Computational Astrophysics and Department of Astronomy \& Physics, Saint Mary's University, 923 Robie Street, Halifax, NS B3H 3C3, Canada}

\author[0000-0003-1227-3084]{Michael R. Meyer}
\affil{Astronomy Department, University of Michigan, Ann Arbor, MI 48109, USA}

\author[0000-0002-8512-1404]{Takahiro Morishita}
\affil{Space Telescope Science Institute, 3700 San Martin Drive, Baltimore, MD 21218, USA}

\author[0000-0002-8530-9765]{Lamiya Mowla}
\affiliation{Dunlap Institute for Astronomy and Astrophysics, University of Toronto, 50 St George Street, Toronto, ON M5S 3H4, Canada}

\author[0000-0002-9330-9108]{Adam Muzzin}
\affil{Department of Physics and Astronomy, York University, 4700, Keele Street, Toronto, ON MJ3 1P3, Canada}

\author{Ga\"el Noirot}
\affil{Institute for Computational Astrophysics and Department of Astronomy \& Physics, Saint Mary's University, 923 Robie Street, Halifax, NS B3H 3C3, Canada}

\author[0000-0003-4196-0617]{Camilla Pacifici}
\affil{Space Telescope Science Institute, 3700 San Martin Drive, Baltimore, MD 21218, USA}

\author[0000-0002-1715-7069]{Neil Rowlands}
\affil{Honeywell Aerospace, 303 Terry Fox Dr, Ottawa, ON K2K 3J1, Canada}

\author{Ghassan Sarrouh}
\affil{Department of Physics and Astronomy, York University, 4700, Keele Street, Toronto, ON MJ3 1P3, Canada}

\author[0000-0002-7712-7857]{Marcin Sawicki}
\affil{Institute for Computational Astrophysics and Department of Astronomy \& Physics, Saint Mary's University, 923 Robie Street, Halifax, NS B3H 3C3, Canada}

\author[0000-0003-4068-5545]{Joanna M. Taylor}
\affil{Space Telescope Science Institute, 3700 San Martin Drive, Baltimore, MD 21218, USA}

\author[0000-0002-3824-8832]{Kevin Volk}
\affil{Space Telescope Science Institute, 3700 San Martin Drive, Baltimore, MD 21218, USA}

\author[0000-0002-9842-6354]{Johannes Zabl}
\affil{Institute for Computational Astrophysics and Department of Astronomy \& Physics, Saint Mary's University, 923 Robie Street, Halifax, NS B3H 3C3, Canada}

\begin{abstract}
We present the wide field slitless spectroscopy mode of the NIRISS instrument on the {\it James Webb Space Telescope}. This mode employs two orthogonal low-resolution (resolving power $\approx 150$) grisms in combination with a set of six blocking filters in the wavelength range 0.8 to $2.3\,\mu$m to provide a spectrum of almost every source across the field-of-view. When combined with the low background, high sensitivity and high spatial resolution afforded by the telescope, this mode will enable unprecedented studies of the structure and evolution of distant galaxies. We describe the performance of the as-built hardware relevant to this mode and expected imaging and spectroscopic sensitivity. We discuss operational and calibration procedures to obtain the highest quality data. As examples of the observing mode usage, we present details of two planned Guaranteed Time Observations programs: The Canadian NIRISS Unbiased Cluster Survey (CANUCS) and The NIRISS Survey for Young Brown Dwarfs and Rogue Planets.
\end{abstract}


\section{Introduction}

The \jwst\ (JWST) is a 6.5\,m-aperture telescope that was launched on 25 December 2021 to a halo orbit around the second Sun-Earth Lagrange point. Equipped with a suite of near-infrared (NIR) to mid-infrared (MIR) instruments it will be a revolutionary facility for furthering humanity's understanding of the cosmos. Some of the key features of JWST are the large collecting area, high spatial resolution, low background from cooled telescope and instruments and very stable image quality. The telescope was designed to be a general purpose observatory with instrumentation offering a large variety of observing modes to target objects from within our own solar system to the furthest known galaxies in the early universe.

Within the wavelength range of 0.6 to 5.3\,$\mu$m covered by the three NIR instruments, JWST offers a total of eight spectroscopic modes. The Near Infrared Imager and Slitless Spectrograph (NIRISS, Doyon et al. in preparation; hereafter Paper I) has a single object grism spectrum mode (SOSS, Albert et al. in preparation) for individual bright objects and a wide field slitless spectroscopy mode (WFSS) utilizing two low-resolution ($R\approx 150$) grisms that is the focus of this paper. The Near Infrared Camera (NIRCam) has two grisms operating at longer wavelengths than those of NIRISS for both wide field and single object slitless spectroscopy \citep{Greene:2017}. The Near Infrared Spectrograph (NIRSpec, \citealt{Jakobsen:2022}) has the first space-based configurable multi-object spectrograph (MOS), a $3\asec$ IFU, several fixed slits and a $1.6\asec$ square aperture for time-series observations. For multi-object observations at 0.8 to 2.3\,$\mu$m the two options are NIRISS WFSS and NIRSpec MOS. In this article we describe in detail the capabilities and usage of NIRISS WFSS and where appropriate contrast this with the alternative capability of NIRSpec MOS. For a complete description of NIRSpec MOS we refer the reader to \cite{Ferruit:2022}.

In slitless spectroscopy, light from the full field passes through a dispersive element such as a grism and a spectrum of almost every source within the field-of-view falls onto the detector. It is therefore a mode that provides a very high multiplex gain with potentially thousands of spectra recorded simultaneously. In such a slitless mode the full field sky background emission is also
dispersed by the grisms with the result that there is a dispersed spectral
background that is effectively the mean spectrum from many background points.
Hence the spectral background is about the same as the imaging
background and is higher than can be obtained in a configurable MOS.

With slitless spectroscopy there is no flux excluded from a finite slit width, resulting in no PSF-dependent `slit losses', accurate flux calibration and time-series observations free of time-variable slit loss. For low spectral-resolution observations, careful subtraction of the continuum can isolate maps of line emission, making this mode especially useful for studies requiring accurate emission line structure and/or ratios \citep{Brammer:2012a}. With the excellent angular resolution of the {\it James Webb Space Telescope}, scales of a few hundred parsecs in galaxies at redshift $z=2$ can be mapped to study their structural evolution. 

Another benefit of slitless spectroscopy is that observations are quite simple to execute. There is no need for pre-imaging of the field or accurate astrometry of potential spectroscopic targets. No target acquisition is required to place the targets within small apertures as the telescope pointing uncertainty is a thousand times smaller than the field of view. These features make wide field slitless spectroscopy a powerful mode for parallel observations. With parallel observations, extra data can be obtained  
`for free' whilst another instrument is performing a prime observation. This enables large survey programs covering an area that would be prohibitive to perform with prime observations. It is therefore not surprising that NIRISS WFSS has been allocated the majority of the pure parallel observing time during the first year of JWST operations\footnote{\url{https://www.stsci.edu/jwst/science-execution/approved-programs/cycle-1-go}}.

Wide field slitless spectroscopy has been available on the \hst\ (HST) using the ACS G800L, WFC3/UVIS G280, WFC3/IR G102 and G141 grisms. Scientific highlights of the usage of these HST grisms include spectral template definition, binarity and variability of the coolest brown dwarfs \citep{Schneider:2015,Buenzli:2015}, detailed characterization of gravitationally-lensed star-forming galaxies \citep{Brammer:2012b,Wang:2020}, spectroscopic confirmation of  $z\sim 2$ clusters \citep{Noirot:2018}, age-dating passive galaxies in $z\sim 2$ clusters \citep{Newman:2014} and in the field \citep{Whitaker:2013,Estrada-Carpenter:2019,Morishita:2019}, discovery of extreme emission line galaxies \citep{Atek:2011}, statistical studies of galaxy evolution \citep{Whitaker:2014,Prusinski:2021}, discovering the prevalence of inside-out star formation \citep{Nelson:2016},  measuring galaxy properties, including \lya\ absorption by the intergalactic medium, in the reionization epoch \citep{Schmidt:2016,Tilvi:2016} and breaking the record for most distant spectroscopic redshift of a galaxy at $z=11$ \citep{Oesch:2016}. The 3D-HST survey alone measured $\sim$100,000 spectroscopic redshifts with the WFC3/IR grisms \citep{Momcheva:2016}, vastly more than all ground-based near-IR spectroscopy combined. Given the wealth of science done with HST slitless grisms, it can be expected that NIRISS on board JWST will prove similarly productive.

\begin{figure*}
\begin{center}
\includegraphics[scale=.85]{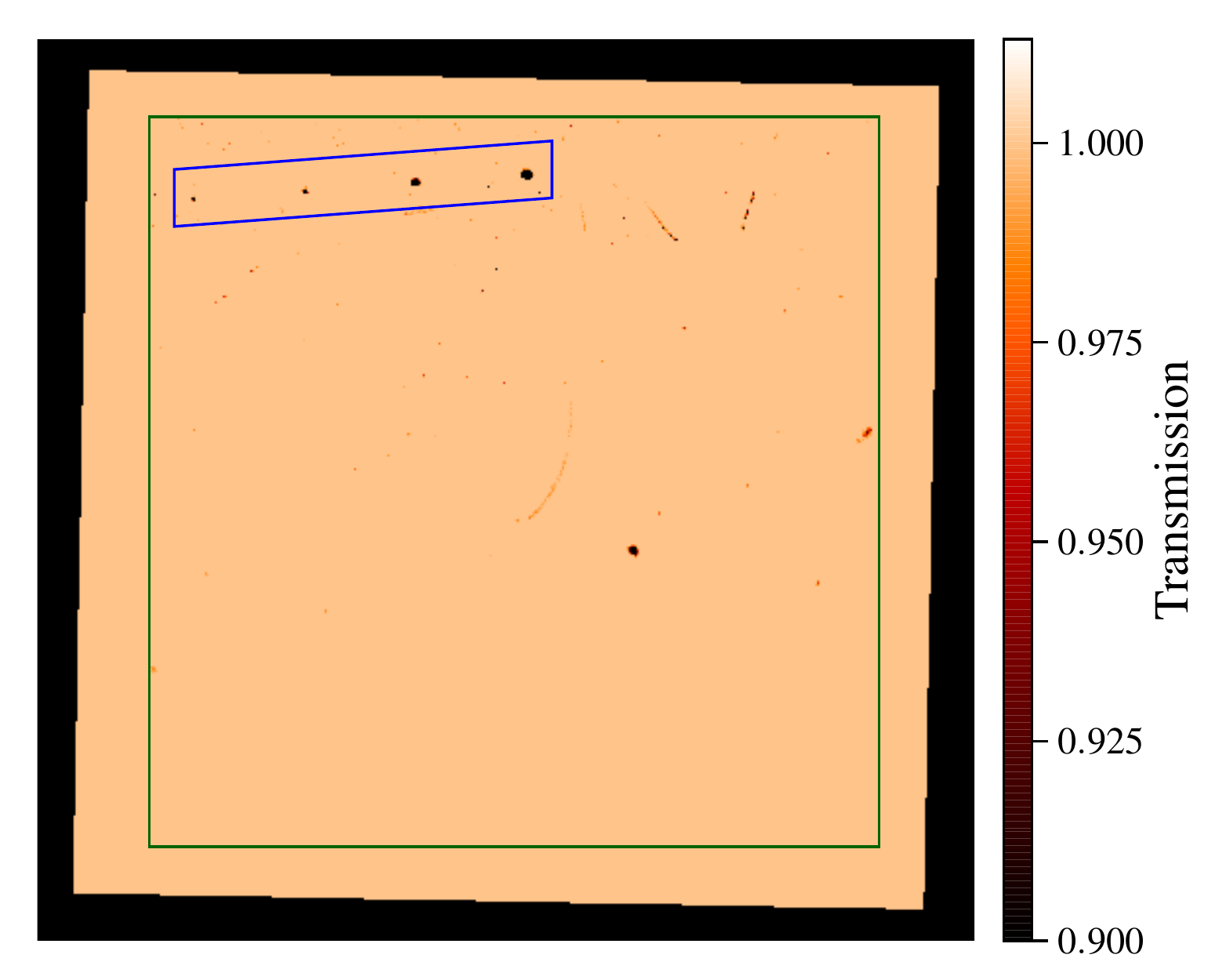}
\caption{Optical transmission of the NIRISS pick-off mirror. The black border shows part of the region beyond the mirror edges. The outline of the region of the POM that images onto the detector is displayed with a green box. The four coronagraphic spots lie within the blue parallelogram. Numerous small features are visible, particularly in the upper region that is in focus. However, most of these features decrease transmission by only a few percent.
\label{fig:pomtrans}}
\end{center}
\end{figure*}

This paper is one of a series of five describing the NIRISS instrument, its three main observing modes, and the Fine Guidance Sensor. In Section \ref{sec:hardware} we describe aspects of the as-built hardware important for wide field slitless spectroscopy and in Section \ref{sec:performance} the expected on-sky performance based on ground-test data. In Section \ref{sec:opscon} we discuss the operations concept including calibration. Section \ref{sec:dataproc} details the data processing and analysis steps. In Section \ref{sec:gto} we present two Guaranteed Time Observations (GTO) programs that illustrate the science use of wide field slitless spectroscopy. Section \ref{sec:summary} provides a summary. All figures of the NIRISS detector in this paper use the Data Management System (DMS) orientation, such that the spacecraft +V3 axis is up and the +V2 axis to the left. There is a $0.6^\circ$ rotation between the V3-V2 axis and the NIRISS detector axis.
Further information on NIRISS is available on the JWST User Documentation\footnote{\url{https://jwst-docs.stsci.edu}} site hosted by the Space Telescope Science Institute. 

\section{NIRISS Hardware}\label{sec:hardware}

\subsection{Overview}

NIRISS is a fully reflective instrument containing gold-coated mirrors for high infrared throughput. It takes light from the pick-off mirror, passes it through a collimator three-mirror anastigmat, then through a pupil wheel and filter wheel. Next a camera three-mirror anastigmat focuses the light on to the detector. The different imaging or spectroscopic observing modes are selected only by rotating the pupil wheel and/or filter wheel to pass the light through filters, grisms, the non-redundant mask and clear apertures. 

Originally, the instrument was named the Tunable Filter Imager (TFI) and was based around a Fabry-Perot etalon, enabling tunable narrow-band imaging. However, to avoid schedule risk associated with this new technology, TFI was converted into NIRISS in 2011 \citep{Haley:2012}. The etalon and its order-limiting filters were removed from the design and three grisms and a range of wide- and medium-band filters were added to the pupil and filter wheels to enable the NIRISS observing modes.
NIRISS was tested in a series of cryo-vacuum campaigns first at the
David Florida Laboratory in Canada and later at the Goddard Space Flight
Center and the Johnson Space Center in the USA between 2011 and 2017. Most results shown in this paper are from the third and last Goddard cryo-vacuum test known as CV3 \citep{Kimble:2016}.

\subsection{Pick-off Mirror}

The Pick-off Mirror (POM) feeds the light from the telescope into the instrument. The POM transmission mask derived from ground test data is shown in Figure \ref{fig:pomtrans}. There are a few characteristics of the POM that are particularly relevant for direct imaging and wide field slitless spectroscopy. The POM is oversized compared to the detector to ensure that the detector area is fully utilized. Therefore when a grism is inserted into the beam it is possible for spectra from sources outside the imaging field of view to fall on the detector. 
Additionally, the astrophysical background (see Section \ref{sec:background}) illuminating the POM will be dispersed onto the detector leading to a distinct series of bands in the structure of the background corresponding to the filter cutoffs of various spectral orders. 

The original TFI design of the instrument incorporated a coronagraphic imaging mode for high-contrast imaging. Four occulting spots with diameters $0.58\asec, 0.75\asec, 1.5\asec, 2.0\asec$ were engraved into the POM. The spots reduce the flux for any object and the background in those field locations. The effect of the coronagraphic occulting spots will be clearly visible in NIRISS WFSS direct images and spectra. Because of the relatively small dither patterns used in NIRISS WFSS there will usually remain missing data at the locations of the larger spots, even after dithering and combining exposures. 

Several small spots and linear or curved features that decrease transmission have been identified using ratios of dispersed and direct imaging flats that eliminate quantum efficiency and detector defects from the resulting ratio image (Figure  \ref{fig:pomtrans}). Most of these features decrease transmission by up to 10\%, although the largest feature to the lower-right of center blocks approximately 50\%. The transmission is wavelength dependent with higher transmission at longer wavelengths. Because of a designed tilt in NIRISS, only the upper part of the POM is at focus on the detector, so such features are more readily identified in the upper half of the image and more diffuse in the lower half. These features are either surface defects in the gold coating or contamination. They have been stable across multiple cryo-cycles, however an expected shift in their position on the detector for changes in focus has been observed. After NIRISS achieves best focus during commissioning, their location and transmission
are not expected to change significantly during the JWST mission.

\subsection{Detector}

NIRISS employs a single Teledyne HAWAII-2RG chip (H2RG; \citealt{Loose:2003}) as its detector. It is sensitive to light with wavelengths from 0.6 to 5.3\,$\mu$m, with quantum efficiency greater than 80\% over most of that range. The detector has an array of $2048\times2048$ pixels of which an outer perimeter of 4 pixels are reference pixels, not sensitive to light, leaving $2040\times2040$ active pixels. With a scale of $0.065 \asec$ per pixel, the imaging and WFSS field-of-view is $2.2\amin \times 2.2\amin$. There is a rotation of 1.2$^\circ$ between the alignment of the POM and the detector. We briefly describe relevant aspects of the detector performance here while further detail can be found in Paper I.

The H2RG detector is read out using non-destructive up-the-ramp sampling, with one frame every 10.7 seconds in full-frame mode. There are two readout modes available: NISRAPID and NIS. For NISRAPID, each frame is transmitted to the ground, whereas for NIS, the onboard computer averages every 4 frames together into a single group and only each group is transmitted to the ground. The noise in data collected with NIS readout is expected to perform similarly to NISRAPID, with the main difference being slightly worse performance in cosmic ray identification, particularly when a cosmic ray impact occurs in the middle of a group.

The selected detector has low dark signal of 0.39 electrons per frame (equivalent to full-frame dark current of 0.036 e-/s) and low total noise in full-frame-readout ramps (8.4 e- for a 1000\,s ramp). For short ramps the $1/f$ noise is a major noise component and leaves significant stripes in the calibrated (reference-pixel corrected) data. In imaging mode it is possible to correct for some of this striping by masking objects and subtracting the median of each column. The striping is more concerning for spectroscopy because the spectral traces are approximately aligned with the detector rows and columns when using the GR150C and GR150R grisms, respectively. However, for long spectroscopic integrations that approach 1000\,s and contain a typical zodiacal and scattered light background, the striping will be almost unnoticeable because the noise is dominated by the background. The high cosmic ray rate at L2 will be an additional source of noise that affects the overall sensitivity.

\begin{figure*}
\begin{center}
\includegraphics[scale=.42]{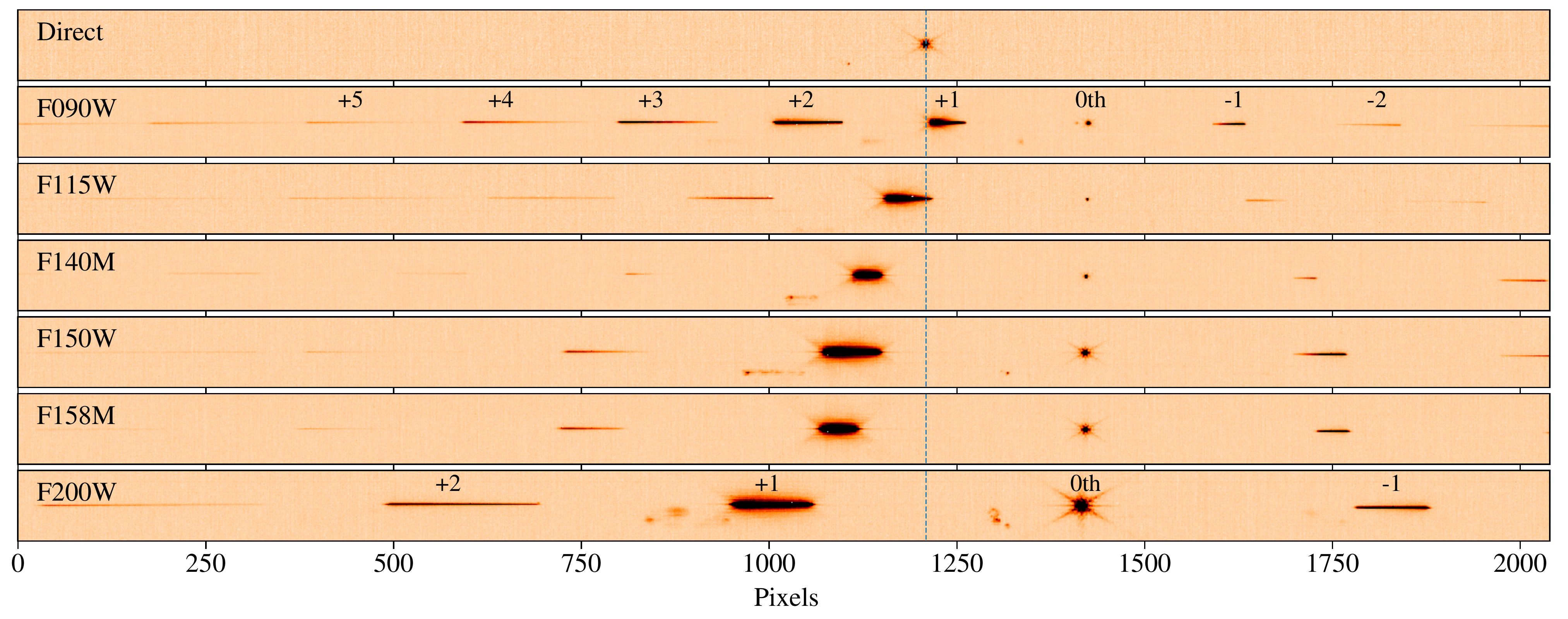}
\end{center}
\caption{Layout of dispersed spectra across the full width of the NIRISS detector compared to the direct imaging source position (top panel, also marked by vertical dashed line in all panels). These ground-test observations used an artificial star source at CV3. Locations of each order are labeled for two of the six blocking filters. The NIRISS first-order spectra in each filter are relatively short at $\sim 100$ pixels in length, resulting in much lower source overlap than with other instruments, such as WFC3/IR and NIRCam. The spectra of the diffuse object to the lower-left of the artificial star is the result of the optical ghosts described in Section \ref{sec:performance}.
\label{fig:dispersedps}}
\end{figure*}

The NIRISS detector has good cosmetics with a low number of bad pixels (0.9\%). In addition to these, a further 2\% of pixels show some excess noise in dark exposures that may or may not be significant depending on the science case. Ground testing showed that the bad pixel population changes with detector temperature. However, it is planned that the NIRISS detector will operate at a constant temperature for all of its lifetime, negating any such temperature-dependent bad pixel changes. Bad pixel maps are available based on ground test data and will be updated regularly after NIRISS reaches its final operating temperature. 


\subsection{Grisms}

 Grism spectra from different objects often overlap, particularly in crowded fields. The excellent sensitivity of NIRISS on JWST means that most fields will be crowded for moderate to long exposure times. To alleviate this problem, NIRISS employs two identical grisms, GR150C and GR150R, oriented with orthogonal dispersion directions approximately aligned with the detector \textit{rows} and \textit{columns}, respectively. Note that the `C' and `R' are reversed because the grisms were named using the original detector orientation, before the switch to the DMS orientation. Having two orthogonal grisms also allows one to triangulate a position and wavelength when a source is only identified by a single emission line. 
 
The two GR150 grisms are blazed at a wavelength of 1.3\,$\mu$m. The grisms have a peak first-order transmission of 80\% and first-order throughput above 50\% between 1.0 and 2.1\,$\mu$m. The second order peaks at 0.68\,$\mu$m. Further details on the measured grism transmission versus wavelength are given in Paper I.

Each grism provides a mean resolving power $(\lambda / \Delta \lambda)$ of 150 (in first order) over the wavelength range 0.8--2.3 $\mu$m. The dispersion is $\sim$ 47 \AA\ pixel$^{-1}$ and two pixels sample one resolution element. The first order spectrum at a wavelength of 1.047 $\mu$m is not deviated by the grism, and falls on approximately the same pixel as in the direct images.  

Ground tests have revealed that the filter-wheel mechanism can place either of the grisms within $\pm 0.15^{\circ}$ of its nominal position.  Offsets of the grism element from the default location in the wheel impart a rotation to the entire field of view.  A trace solution has been developed that depends on the position of the filter wheel determined from an encoder, and can be read from the header keyword \texttt{FWCPOS}. In ground test data there is a good correlation between \texttt{FWCPOS} and source trace rotation angle measured on the detector. During commissioning the trace angles for many observations will be correlated with \texttt{FWCPOS} to determine the relationship and its uncertainty. The data processing software will use \texttt{FWCPOS} to determine the predicted trace location and orientation.

\subsection{Filters}

The WFSS grisms in the NIRISS filter wheel must be crossed with one of the band-limiting filters located in the pupil wheel. These filters restrict the wavelength range of spectra and are one of the keys to mitigating source overlap in very deep NIRISS spectroscopy. The first-order NIRISS WFSS spectra are much shorter than those of the WFC3/IR grisms that do not utilize blocking filters. There are six filters available for WFSS, four are broad-band (F090W, F115W, F150W and F200W) and two are medium-band (F140M and F158M). Five of the filters have the same specifications as filters in NIRCam, the exception being F158M. Paper I tabulates the central wavelength, bandpass edges and average transmission of all NIRISS filters. The ``W'' filter set is based roughly on ground-based near-IR filters and not designed specifically for spectroscopic band-limiting, so there are small gaps in wavelength between the edges of the F115W and F150W filters and the F150W and F200W filters. Thus when using several filters to obtain a spectrum with broad wavelength range there will be small gaps in wavelength coverage and consequently certain lines will be unobservable at certain redshifts (see Section \ref{sec:spectroscopy}). The layout of a dispersed point source using each blocking filter is shown in Figure \ref{fig:dispersedps}.
The shortest wavelength filter, F090W, has a bandpass from 0.8 to 1.0\,$\mu$m. In this filter the second order throughput is higher than the first order between 0.8 and 0.9\,$\mu$m, so at these wavelengths both orders should be utilized for optimum sensitivity.

\section{Performance}\label{sec:performance}

\subsection{Direct Imaging}

To provide accurate wavelength calibration and spectral trace normalization and location, it is important to have a direct image through the same filter, without the grism in the optical path. In Cycle 1 it will be mandatory to obtain NIRISS direct imaging in WFSS mode to ensure all data can be calibrated. It may be possible to use other methods to provide a direct image location such as (i) the zeroth order location, (ii) the intersection of the two grism traces, (iii) imaging obtained with a different filter, (iv) imaging obtained at a different time with NIRISS or NIRCam. However, the quality of the calibration using other methods like these will need to be verified after launch. 

In slitless spectroscopy the spatial resolution of the imaging and spectroscopic data are identical. In the six WFSS blocking filters the FWHM of a point source ranges from 42 (F090W) to 65 (F200W) milli-arcseconds (mas). However, the images will be under-sampled by the detector pixel scale of 65 mas per pixel. The telescope optics are not diffraction-limited at wavelengths less than  2$\mu$m, such that there is a significant halo around the PSF core. The expected 80\% encircled energy radius ranges from $0.229\asec$ (F090W) to $0.168\asec$ (F200W), according to the WebbPSF simulation tool \citep{Perrin:2014}. The PSF will be regularly monitored for changes, but is expected to be very stable. The WebbPSF tool allows users to generate oversampled (sub-pixel) PSFs.

The imaging sensitivity of NIRISS is described in detail in Paper I. Here we are interested in the imaging sensitivity for direct imaging to enable source identification for spectroscopy. A general rule-of-thumb for low-resolution grism spectroscopy is that the total direct imaging exposure time when background-limited should be $\approx 10$\% of the total spectroscopic exposure time. For a spectrum of a continuum-dominated source that extends $\approx 100$ pixels in the spectral direction (compared to $\approx 2$ pixels for a direct image), in the background-limited noise regime, this gives an imaging S/N a factor of $\approx 15$ higher than the grism S/N per spectral pixel. If the targets have photometry dominated by emission lines then the fraction of time for direct imaging may need to be greater than $10$\% to securely detect such objects. Alternatively, emission lines may be identified solely in the spectra by triangulation from their positions with the two orthogonal grisms.

\begin{figure}
\includegraphics[scale=.42]{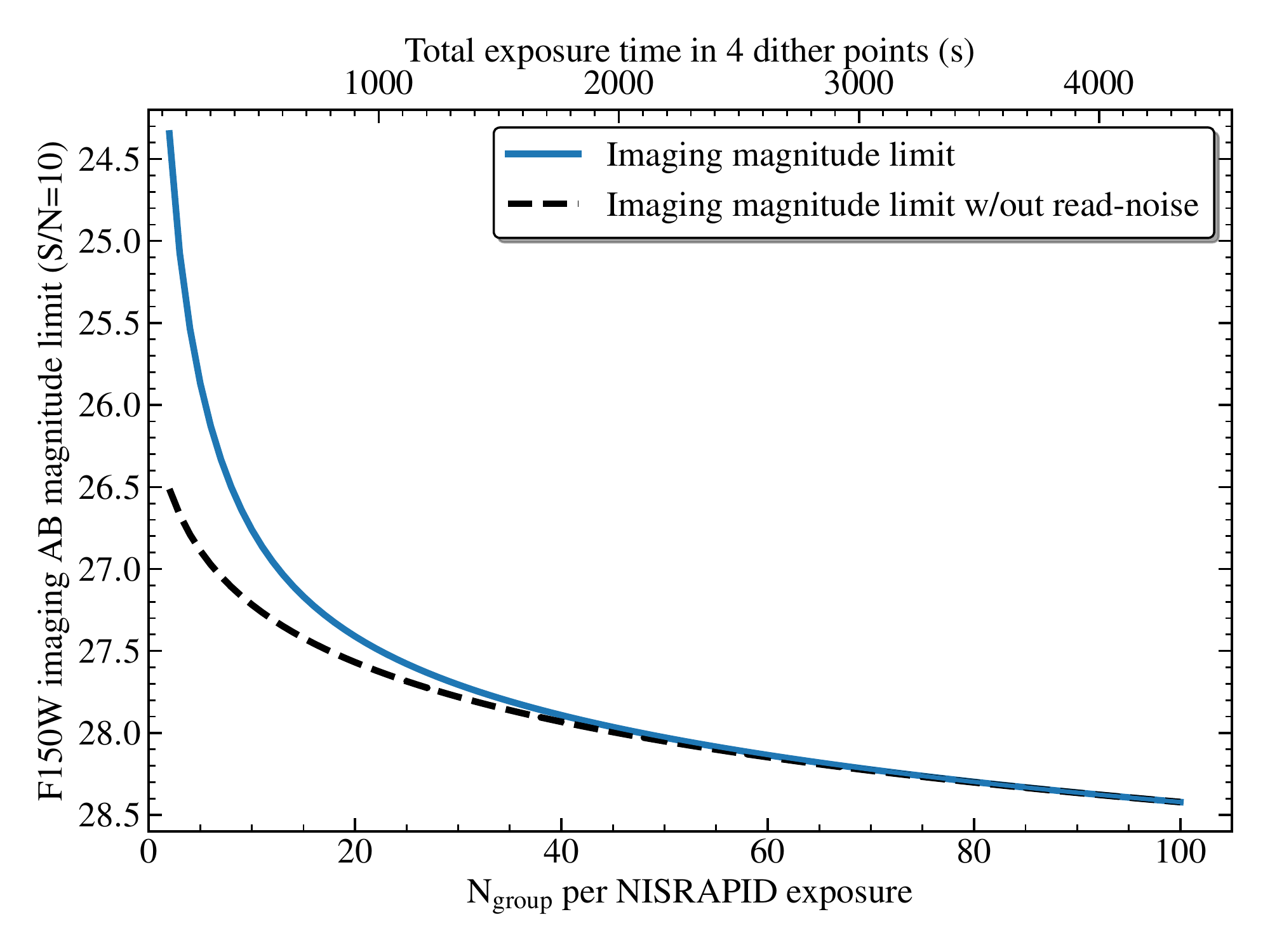}
\caption{Point-source limiting magnitude in imaging mode with the F150W filter as a function of number of groups per exposure in NISRAPID readout mode for a 4 point dither pattern of direct imaging (blue). The upper axis shows the total exposure time of the 4 exposures. The dashed black line shows the scaling to shorter times if the exposures were background limited (setting detector read-noise to zero). The dominant contribution of detector read-noise for short exposures ($\rm{N_{group}}<20$) leads to a relatively bright magnitude limit.   
\label{fig:directsnr}}
\end{figure}

For the case of shallow grism spectroscopy with few dithers and short exposure times, the 10\% rule for direct imaging may become extremely inefficient due to a high fraction of time for exposure overheads and read-noise dominance at short exposure times. In Figure \ref{fig:directsnr} we show the F150W filter $10\sigma$ point-source limiting magnitude for the case of four direct image dither positions as a function of the number of groups per exposure and the total exposure time. It can be seen that very short exposures should be avoided because of the imaging depth achieved.

For fields containing bright objects, the user should be careful to avoid or mitigate the effects of saturation in direct imaging. For standard imaging of a single target the usual method to avoid saturation is to utilize a sub-array on the detector that is read out in a fraction of the regular frame time of 10.7\,s. Sub-array imaging is only available for NIRISS calibration observations. For regular science all WFSS imaging and spectroscopy must use the full detector readout. Saturation has two important consequences: increased uncertainty on the centroid of an object and therefore the astrometric/wavelength solution for this object on the corresponding grism exposure and detector persistence that will be seen on the subsequent exposure(s). The full-frame direct image point-source saturation limits for the shortest possible integration, two groups in NISRAPID read-out, range from AB magnitude 17.6 for F140M to 18.4 for F115W, F150W and F200W. For NIS read-out the limits are 1.5 magnitudes fainter. Sources fainter than these limits can have a flux measurement and accurate centroiding can be done for them. Such sources may saturate in later reads in the ramp depending on the exposure length, which can produce persistence in subsequent ramps. However this does not need affect the flux and centroid measurements as these can be extracted only from read-out groups recorded prior to saturation.

The band-limiting filters in NIRISS produce ghosting at the 0.1 to 2\% level. When using NIRISS in imaging mode it will be important to understand where in the field the ghost of each bright source is located, so they are not misidentified as independent, faint objects. Analysis of ground test data has shown that the ghosts appear at predictable positions in the field. However, ghost images of point sources do not appear simply as scaled versions of the PSF. They are distorted, with field- and filter-dependent morphologies that include multiple separate components. Therefore, modeling and subtraction of ghosts will be imperfect, so mitigation by masking is likely to be the preferred option, at least for early operations. During commissioning there will be observations of a very large number of stars across the field that will be used to characterize the ghosts in much more detail than was possible in ground tests, potentially enabling modeling and subtraction in the future.

The NIRISS filters induce a small offset in the location of sources on the detector. For all filters except F200W the offset is less than 1 pixel. For the F200W filter the offset is 2.5 pixels ($0.16 \asec$). The astrometric distortion will be measured during commissioning using observations of the JWST astrometric field in the Large Magellanic Cloud (LMC). A separate World Coordinate System (WCS) solution for each filter will be determined.

\subsection{Spectroscopy}\label{sec:spectroscopy}

\begin{figure}
\includegraphics[scale=.42]{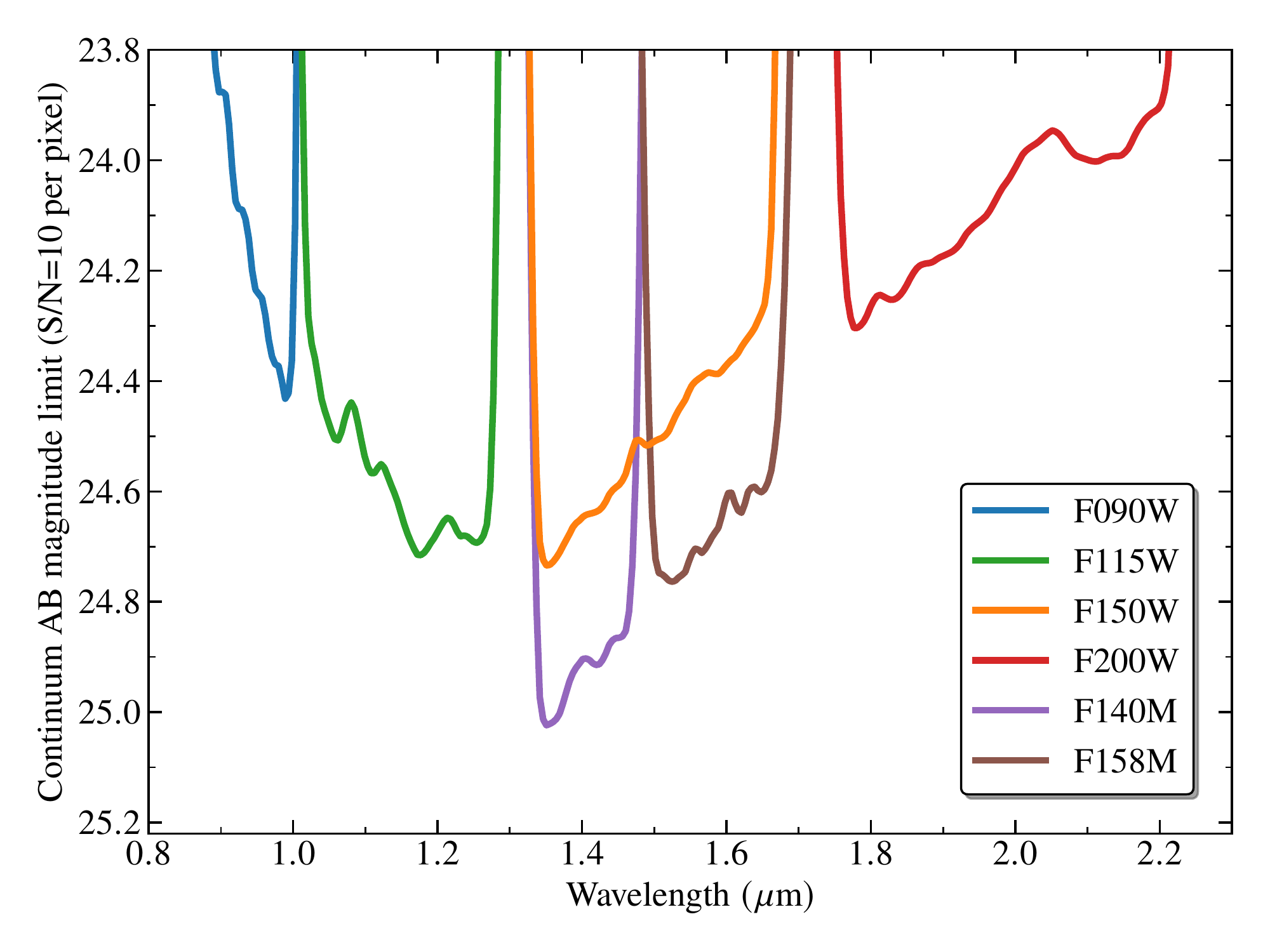}
\caption{Point-source continuum limiting AB magnitude for S/N=10 per pixel in WFSS mode with the GR150R or GR150C grism in combination with each of the six blocking filters. The total exposure time per filter is 10,000\,s. The spectrum is extracted in a box with cross-dispersion aperture of $0.30\asec$. For a resolved source, $n=1$ circular Sersic profile with effective radius of $0.2\asec$ or $0.4\asec$, a similar S/N is achieved for a source 0.4 or 0.8 magnitudes brighter, respectively. In these cases the optimum extraction aperture is correspondingly larger. The limiting magnitudes here only consider the first order spectrum. The second order has higher transmission below 0.9\,$\mu$m. 
\label{fig:grismsnr}}
\end{figure}

\begin{figure}
\includegraphics[scale=.42]{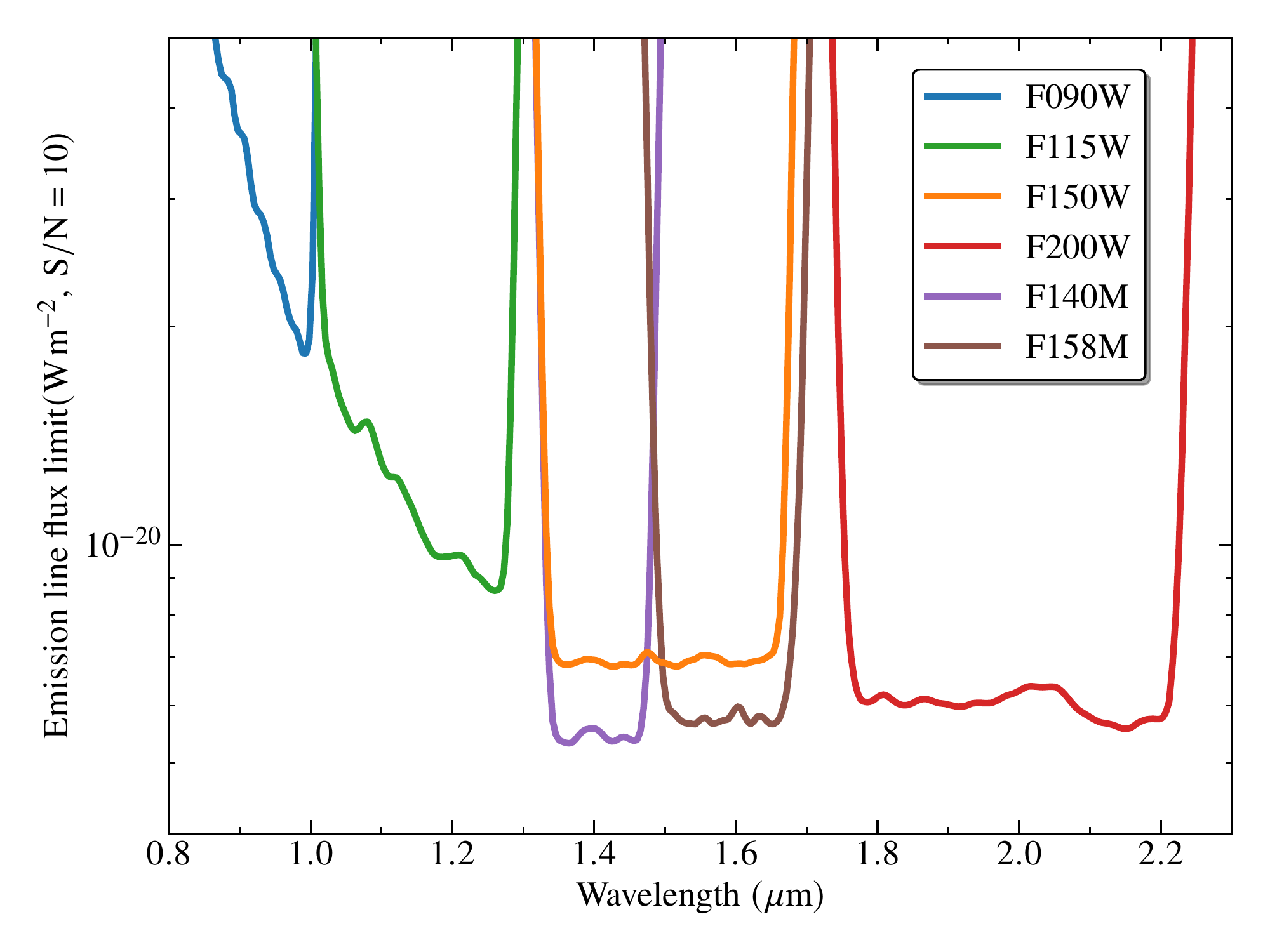}
\caption{Point-source limiting emission line flux for S/N=10 per pixel in WFSS mode with the GR150R or GR150C grism in combination with each of the six blocking filters. The total exposure time per filter is 10,000\,s. The emission line is extracted in a $0.3 \asec$ diameter circular aperture.
\label{fig:grismlinesnr}}
\end{figure}

Figure \ref{fig:grismsnr} shows the $10\sigma$ per spectral pixel point-source limiting AB magnitude in 10,000\,s of background-limited grism spectroscopy for all six blocking filters. The limiting magnitude ranges from 24 to 25 across most of the wavelength range. Figure \ref{fig:grismlinesnr} shows a similar plot for emission line sensitivity. The two Medium filters, F140M and F158M, have the highest sensitivity due to reduced background. Each of these Medium filters requires only half as long an integration time as the Wide F150W filter to reach the same sensitivity. Therefore in a fixed observing time, the same sensitivity is reached using either both Medium or one Wide filter. The Wide filter has continuous coverage without potential bandpass edge effects, whereas the Medium filters have shorter spectra giving less spectral overlap.

The full-frame GR150 point-source saturation limits for NIRISS in the six filters used for WFSS range from AB magnitude 13.3 for F200W to 14.1 for F115W. Brighter sources will saturate a large number of pixels in the first-order spectrum, potentially leading to persistence in subsequent integrations. It is advisable to avoid placing bright sources within the field, and just outside the field but within the POM area where spectra can be dispersed onto the detector.

The ghost of each bright source due to the NIRISS filters described in Section 3.1 also cause dispersed ghosts in the grism data (the spectral orders of the ghost are visible below the main traces in Figure \ref{fig:dispersedps}). Further in-flight data is required to determine whether the ghost spectra are achromatic within each filter. Since the flux in ghosts is a small percentage of the flux in sources, the ghosts should not cause a significant increase in spectral contamination, but may be somewhat more complex to model and subtract. 

\begin{figure}
\includegraphics[scale=.43]{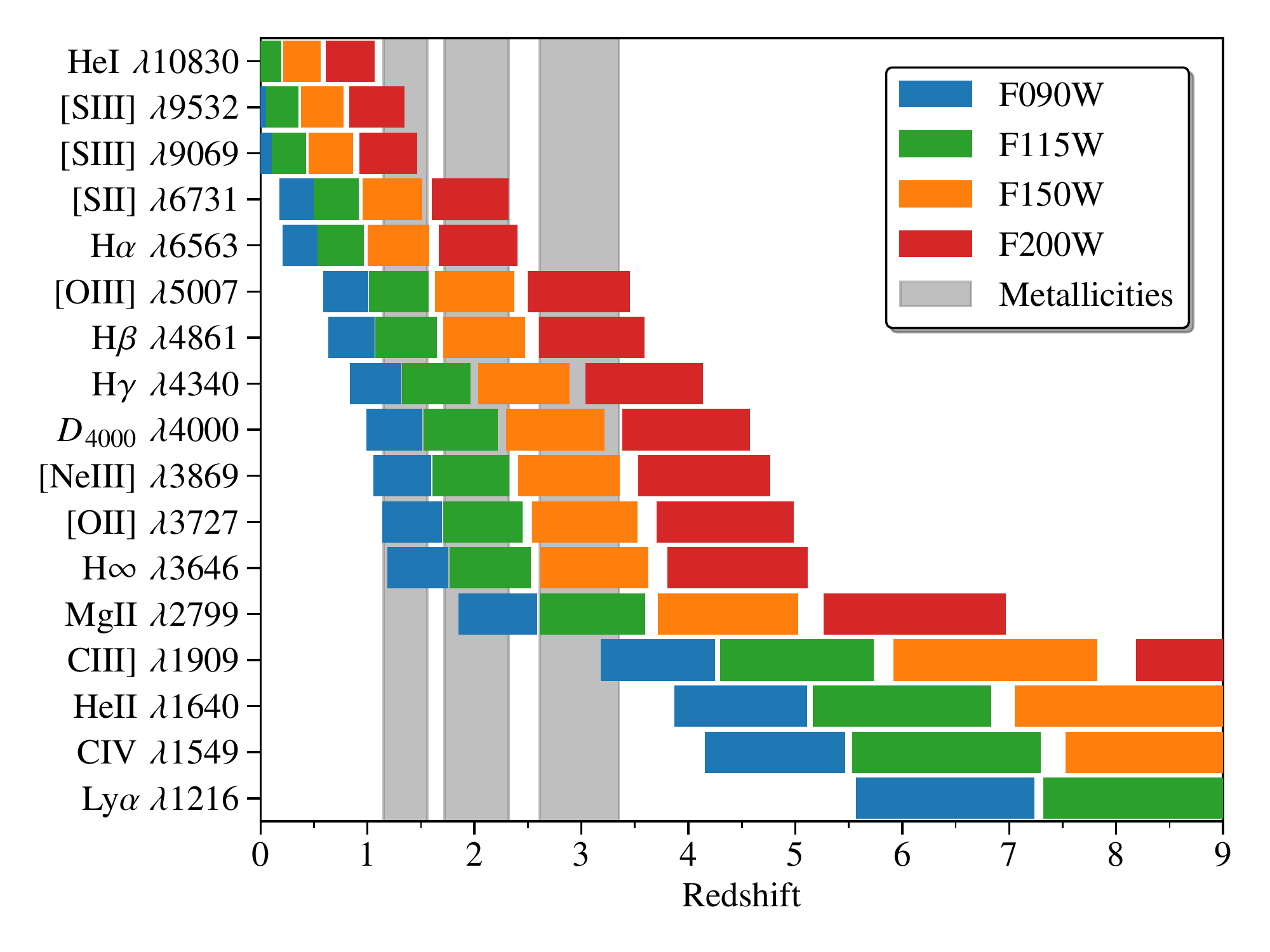}
\caption{Spectral features of galaxies in NIRISS WFSS `W' filters as a function of redshift. For each of the four filters the range of redshift where each feature is in the bandpass is plotted as a rectangle. NIRISS observations with all four filters will contain at least one strong line at most redshifts. The gray bands show redshift ranges where NIRISS WFSS covers all the metallicity-sensitive lines \oiii, \hbeta, \neiii\ and \oii. 
\label{fig:linesvsz}}
\end{figure}

Figure \ref{fig:linesvsz} shows the emission lines and spectral breaks that lie in the NIRISS ``W'' filters for redshifted galaxies. At redshifts up to $z=3.5$ there are multiple strong emission lines visible such as \halpha, \oiii, \hbeta, or \oii. However the effect of the gaps between the F115W/F150W and F150W/F200W combinations are evident as narrow redshift ranges where certain lines are not observable with NIRISS. Unfortunately, the gaps for the two strongest lines in high-$z$ galaxies (\halpha\ and \oiii) line up for a small range in redshift around $z=1.6$. \lya\ enters the WFSS wavelength range at $z>5.6$ and would be visible at most redshifts up to $z=17$, whenever the line is transmitted through the intergalactic medium. The gray bands show redshift ranges where all of the lines \oiii, \hbeta, \neiii\ and \oii\ are visible within NIRISS WFSS filters. These strong lines, particularly the flux ratios \oiii/\oii\ and \neiii/\oii, can be used to measure high-redshift gas-phase metallicities \citep{Sanders:2020}. The \halpha\ and \hbeta\ lines can both be observed between $z=0.7$ and $z=2.4$ to determine the spatial distribution of star formation and nebular dust extinction.

\section{Operations concept and calibration}\label{sec:opscon}

\begin{figure*}
\hspace{-0.2cm}
\includegraphics[scale=.67]{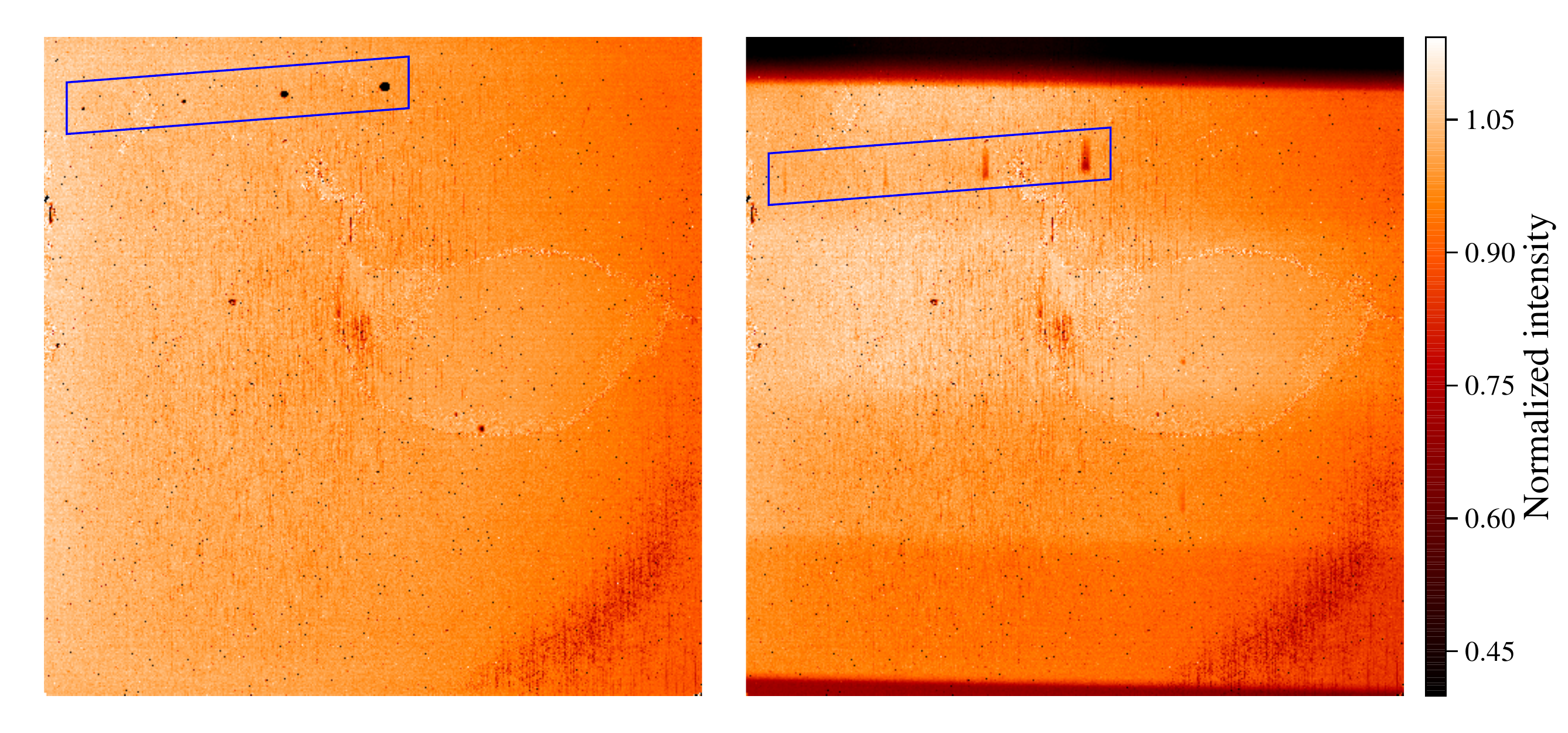}
\caption{Normalized NIRISS flat-field images from CV3 testing. The left side is an imaging flat with F200W+CLEAR and the right side is a dispersed flat with F200W+GR150R. The decreased flux regions from the four coronagraphic spots lie within the blue parallelograms. These regions are seen as dark circles in the imaging flat and negative first order spectra in the dispersed flat. The horizontal bands in the dispersed flat are due to various spectral orders being cut-off by the pick-off mirror, with the two strongest being the first-order at the top of the detector and the zeroth-order at the bottom. The slight tilt of the bands is due to the small rotation between the detector and the pick-off mirror.
\label{fig:flatpom}}
\end{figure*}

\subsection{Observing Sequence and Dithers}
As mentioned in Section 3.1, direct images are taken along with the grism exposures to enable proper identification of the sources in the dispersed images, and also to determine the absolute wavelength zeropoint for the spectra. The NIRISS WFSS observing sequence in any desired filter consists of a direct image (pre-imaging), followed by dithered grism exposures and then a direct image (post-imaging) at the end of the sequence. Direct images are obtained at both the start and the end of the sequence to enable generation of a cleaner combined image by masking of bad pixels. The grism exposures can use either one of the grisms or use the two grisms in succession by selecting the {\it BOTH} option in the Astronomer's Proposal Tool (APT). If multiple filters are required for wavelength coverage, then a new filter is selected in the pupil wheel and the entire sequence is repeated. The WFSS observing sequence is designed to avoid any mechanism moves across dithers, which is important when this mode is used for parallel observations.

NIRISS observing modes use dithers to mitigate detector artifacts, which include bad pixels, persistence, flat-field errors, and pixel sensitivity variations. Most importantly, dithering is necessary for the WFSS mode to improve the sampling of the Point Spread Function (PSF). The NIRISS detector has pixel sizes that critically sample the PSF at $\sim$ 4 $\mu$m, and the PSF is heavily under-sampled at the wavelengths (0.8$-$2.3 $\mu$m) where the WFSS mode operates. The WFSS observations are dithered by selecting from a set of pre-defined dither patterns available in the APT. The dither patterns are defined such that the large pixel offsets allow to mitigate the detector artifacts, and the fractional pixel offsets are used to recover the PSF sampling and structural information in the direct and dispersed images. The WFSS dithers are specified by the {\it number of dither positions} (2, 3, 4, 6, 8, 12 or 16 steps) and the {\it amplitude} of the dither step (Small, Medium, and Large). The Small pattern uses a dither step size of $\sim$ 0.3 arcsec, and offers a higher conservation of pixel phase sampling in the presence of varying geometric distortion. This option with small step size is useful for sparse fields of point (or compact) sources. The Medium and Large dither patterns use step sizes of $\sim$ 0.6 and $\sim$1.2 arcsec respectively, and allow to step over larger features. The Medium pattern is appropriate for extragalactic sources at moderate and high redshifts, while the Large pattern is better-suited for large extended sources in an astronomical scene, such as galaxies at low redshifts. 

The choice of the number of WFSS dithers to use depends on the science program and total observation time. As noted in Section 3, short exposures are inefficient due to higher read noise and overheads. Individual exposures should normally be in the range 400 to 1300 seconds long, with the upper limit due to the cosmic ray hit rate to be confirmed during commissioning. If astrometric and photometric precision are important for the science, more dithers sampling different locations at the sub-pixel scale and different pixels should be considered.

As illustrated in Figure \ref{fig:dispersedps}, spectra obtained in filters F140M, F150W, F158M and F200W are significantly offset from their direct image locations. The consequence is that the spectra of some sources in the direct image will fall off the detector and the spectra of some sources outside the direct image but within the POM region will fall on the detector. This would result in only 76\% of sources in the direct image having complete spectra with both grisms and all four W filters on the detector. To improve the overlap between imaged sources and spectra, a grism- and filter-dependent offset observing sequence is used for the case of the {\it BOTH} grisms option with the F140M, F150W, F158M or F200W filters in the APT. By offsetting the field between the GR150R and GR150C sequences a larger field is imaged, albeit at lower depth at the edges with half coverage. This larger imaged field provides a greater overlap so 89\% of sources within a NIRISS-size field have full spectra and images. If the user does not wish to use these offsets, then each grism should be selected separately rather than selecting the {\it BOTH} grisms option.

\begin{figure*}
\includegraphics[]{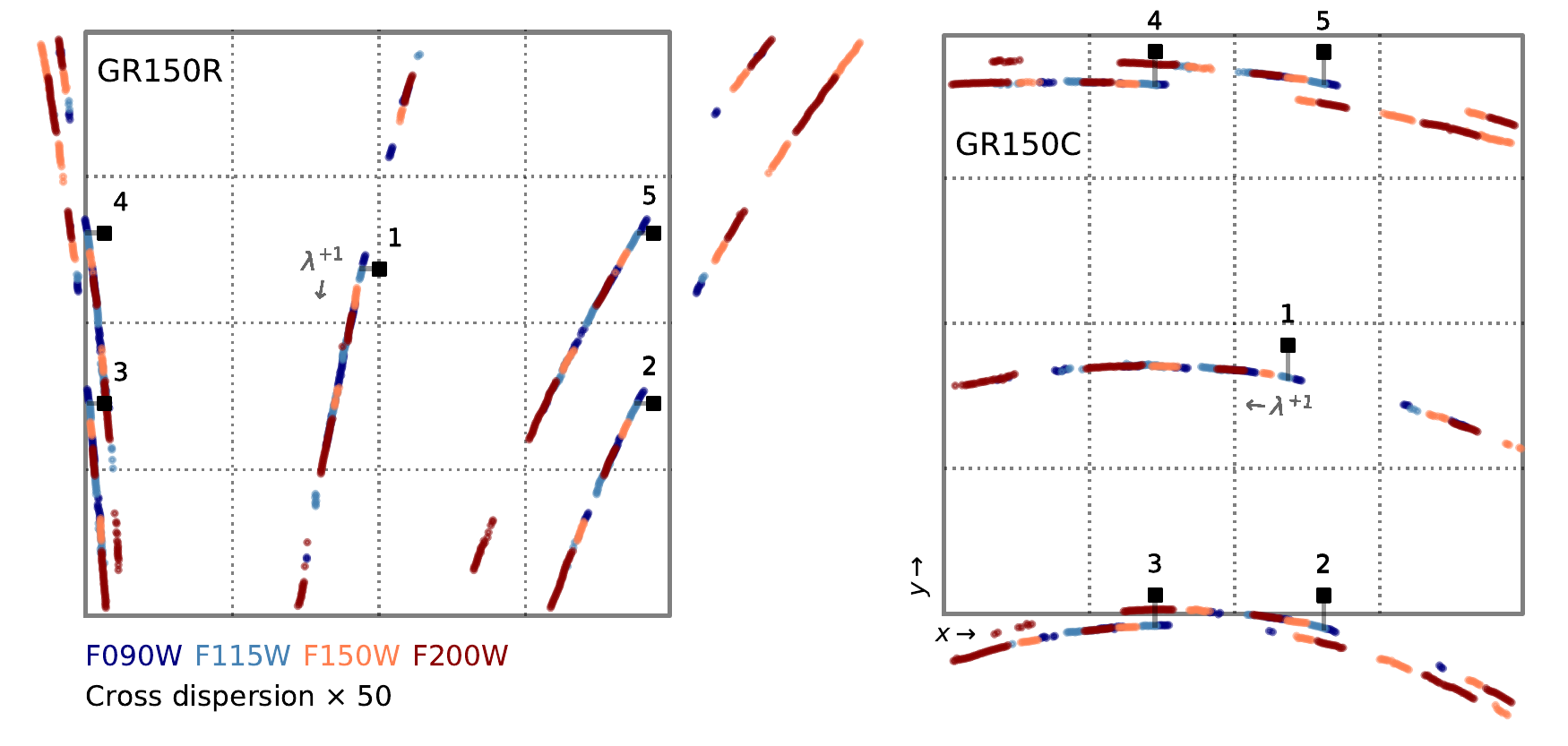}
\caption{Measurements of multiple orders of the spectral traces taken during CV3. The five direct image positions of the calibration source per grism are indicated with black filled squares. They are connected to their associated trace by a small line segment. The cross dispersion scales of the traces and offsets from the direct images have been increased by a factor of 50 for this figure to amplify the small curvature. The grey square in each panel is the detector outline. Although it may appear on this figure that some measured traces fall off the detector, this is just an artifact of the factor of 50 increase in cross-dispersion scale. $\lambda^{+1}$ indicates the first order with the arrow in the direction of increasing wavelength.
\label{fig:trace}}
\end{figure*}

\subsection{Background Subtraction}\label{sec:background}

The major sources of NIR background for JWST are in-field zodiacal emission and stray light. Stray light consists of both zodiacal light and galactic starlight scattered into the field through rogue paths \citep{Lightsey:2016}. It is unknown whether there will be significant spatial structure in the background across the NIRISS field (apart from that caused by effects that can be corrected by flat-fielding) and whether this structure will vary with pointing location. When a grism is inserted the background is dispersed across the field and will show similar banding structure to the flat-field images in Figure \ref{fig:flatpom}. 

The high source density in NIRISS WFSS mode results in a high probability that pixels near a science target trace will be contaminated by flux from other sources. Therefore, one cannot use dedicated regions of the image for each target to model the background and subtract it from the target. Instead the process will be to generate a master background image from the combination of many science images, masking out individual sources. The background level for each exposure will be determined by measurement in source-free regions and the master background scaled to this value before subtraction. This process has been shown to be effective for WFSS with the WFC3/IR instrument. Its accuracy for NIRISS will depend on the variation in background structure at different pointings.

\subsection{Flux Calibration}\label{sec: fluxcal}

Due to the slitless nature of the WFSS mode, there are no slit losses and flux calibration is relatively simple. Observations will be obtained during commissioning and regular science operations of standard stars in imaging mode and WFSS mode using all six filters to determine the efficiency of the various orders of the grisms as a function of wavelength. Any field-dependence of the spectral flux calibration (beyond that taken care of by flat-fielding) will be checked for during commissioning by observing a field of stars.

\subsection{Trace Calibration}\label{sec:tracecal}

During CV3 testing, trace calibration exposures of a bright continuum source in all six filters and both grisms were obtained.  The target was placed at five locations within the field of view in order to investigate the position dependence of the shape of the trace.  In each individual exposure for the unique combination of grism, filter, and target location, we measured the centroid along the trace by fitting a \cite{Moffat:1969} function along the cross dispersion axis.  The traces in the four ``W'' filters are shown in Figure~\ref{fig:trace}. Each trace can be well reproduced by a single quadratic function across all spectral orders and blocking filters. Residuals of the quadratic fit to the measured trace are $\lesssim 0.1$\,pix across the full width of the detector. The direct image location was found to be offset by a few pixels from the trace.

The shapes of the traces in the two grisms are distinct, with GR150C having more curvature.  The traces vary significantly across the field of view, with a larger effect for GR150R. Note the general misalignment of the GR150R spectra with the detector axes is accounted for by the filter-wheel mechanism offset from its nominal position described in Section 2.3. During instrument commissioning, the trace calibration (shape and field dependence) will be verified and refined at a larger number of field points using stars distributed throughout the NIRISS field.

\subsection{Wavelength Calibration}\label{sec:wavecal}

In CV3 a monochromatic source was observed at a wavelength appropriate for each filter and at the same positions across the field of view used for the trace continuum source shown in Figure \ref{fig:trace}.  With knowledge of the intrinsic wavelength of the source and the geometry of the trace, the wavelength calibration is defined by the offset along the trace between the position of the monochromatic line image in the dispersed image and the position of the source in the undispersed direct image.  The CV3 data are consistent with a linear dispersion of the first order spectra of 46.7~\AA/pix and 46.8~\AA/pix for the  GR150C and GR150R grisms, respectively.  During instrument commissioning the compact planetary nebula LMC~SMP~58 \citep{Sanduleak:1978,Stanghellini:2002} will be observed to verify the wavelength calibration.

\section{Data Processing and Analysis}\label{sec:dataproc}

\subsection{JWST Pipeline Processing}

The JWST Science Calibration Pipeline processes JWST data to remove and calibrate instrumental signatures, combine relevant exposures, and perform basic spectral and imaging source extraction. FITS file headers include information from the instrument, observatory, and spacecraft in addition to the relevant proposal details. The pipeline is modular and can be run by users with options to specify their own reference files and configuration parameters.  

Direct images are processed as standard imaging files. They pass through the first stage of the pipeline, \texttt{calwebb\_detector1}, which processes the data from raw non-destructively read ramps to uncalibrated slope images. First, dead, hot, noisy and saturated pixels are flagged and a superbias frame is subtracted. A reference pixel correction is applied using the border of non light-sensitive pixels at the edge of the detector. A non-linearity correction and optional persistence correction/flagging are applied. The dark signal is then subtracted and cosmic rays are flagged. The final step in \texttt{calwebb\_detector1} is to fit a slope to the reads of each pixel. If an exposure contains multiple integrations, a file with the integrations averaged is also output. The second pipeline stage, \texttt{calwebb\_image2}, adds WCS information to the file, applies a flat-field correction and adds flux calibration information to the file. The third pipeline stage, \texttt{calwebb\_image3}, combines multiple exposures (from all dither positions) for each filter into a single, drizzled \citep{Fruchter:1997} image. A catalog and segmentation image constructed from this combined image are used to identify the position and extent of each detected object.

Dispersed images pass through the same pipeline first stage as direct images, but then move on to a different second stage, \texttt{calwebb\_spec2}.  WCS information, including the computed position of the spectrum for each object in the direct image source catalog, is written to the file. \texttt{calwebb\_spec2} masks regions of the dispersed image that contain spectra, scales a master sky background image (provided as a reference file) to the unmasked regions, and subtracts the master sky background image from the entire field.  A flat-field correction is then applied.  The flat-field reference file for each filter is based on the imaging flat-field, but with the POM transmission features removed. It is assumed the flat field is independent of wavelength within each filter bandpass. The flat-fielded two-dimensional image can be optionally output here in case the user wants to do further analysis on the full field, rather than individual object spectra. The two-dimensional spectrum of each object in the field is then extracted and written to a separate image extension in the output FITS file, along with its WCS information. A simple model for spectral contamination from overlapping sources is constructed for each extracted two-dimensional spectrum and subtracted to get the clean, uncontaminated two-dimensional spectrum of each object. The spectra are then flux calibrated and a one-dimensional spectrum extracted. 

For dispersed images, the third stage of the pipeline is \texttt{calwebb\_spec3}.  For each object, the two-dimensional spectra from multiple exposures are combined into a single rectified image (with dimensions of space and wavelength) using a modified drizzle algorithm.  Spectral combination is performed in a single step to avoid noise propagation through multiple interpolations.  The algorithm employs an iterative sigma-clipping technique to reject any cosmic-ray affected pixels that were not detected by \texttt{calwebb\_detector1}. Products are produced for each source with individual exposures in the extensions. From the drizzled two-dimensional spectrum, a one-dimensional spectrum is extracted.

\subsection{Post-Pipeline Processing and Analysis} 

Whilst the JWST pipeline performs a simple contamination correction and outputs extracted one- and two-dimensional spectra, many users will require more specialized software to analyze grism spectroscopy. This is particularly going to be the case in crowded fields. Given the depth achieved by NIRISS grism spectroscopy in moderate to long observations, even `blank' fields may become crowded and require more advanced treatment. 

\begin{table*}[ht]
\caption{CANUCS Prime Target Fields}
\centering
\begin{tabular}{lcccc}
\hline\hline
Cluster & RA & DEC & Redshift & {\it HST} Survey \\ [0.5ex]
\hline
Abell~370         &   02:39:54.1 & -01:34:34  & 0.375 & Frontier Fields       \\
MACS~J0416.1-2403 &   04:16:09.4 & -24:04:21  & 0.395 & Frontier Fields       \\
MACS~J0417.5-1154 &   04:17:35.1 & -11:54:38  & 0.443 & RELICS    \\
MACS~J1149.6+2223 &   11:49:36.7 & +22:23:53  & 0.543 & Frontier Fields      \\
MACS~J1423.8+2404 &   14:23:47.8 & +24:04:40  & 0.545 & CLASH     \\
\hline
\end{tabular}
\label{table:canucstargets}
\end{table*}

The most effective method to simultaneously disentangle overlapping spectra and determine the true nature of the source spectra is by forward modelling of the scene \citep{Brammer:2012a}. The direct image mosaic for a particular filter provides an excellent prior on the locations and flux-density distributions of the sources responsible for each spectrum. Unlike WFC3/IR on HST, NIRISS utilizes the same blocking filters in spectroscopy as in direct imaging, removing the need to interpolate between several imaging filters. 

The shape of the spectra can be modelled using stellar or galaxy templates and the best fit determined by iteratively minimizing the residuals. For galaxies this can also include determining the redshift, potentially by including prior information from photometry over a wider wavelength range than the grism spectroscopy. After subtraction of the continua of the target of interest and overlapping sources, emission lines may be identifiable to give more precise redshifts. All 2D exposures using both grisms, potentially also observed at a range of telescope orientations, can be fit simultaneously, in order to minimize degeneracy between position and wavelength (e.g. \citealt{Ryan:2018,Pirzkal:2018}), and enable efficient use of all data even when some is contaminated. This process can result in maps of each emission line that are particularly useful for subsequent science analysis.

Several software packages that perform analyses similar to described above are available such as \texttt{grizli}\footnote{\url{https://grizli.readthedocs.io/en/latest/}} and \texttt{pyLINEAR}\footnote{\url{https://github.com/Russell-Ryan/pyLINEAR}}.

\section{Planned Guaranteed Time Observations}\label{sec:gto}

The NIRISS Instrument Science Team will use part of their guaranteed JWST observing time to  highlight two of the many applications of NIRISS WFSS mode observations. In this section we provide a high-level overview of these two programs focusing on how WFSS observations enable unique science.

\subsection{CANUCS: The Canadian NIRISS Unbiased Cluster Survey}
\label{highz}

As described in Section 1, wide-field grism observations with {\it HST} have proved to be remarkably successful for addressing a number of science questions in galaxy formation and evolution. With NIRISS on {\it JWST} the improved performance in wavelength range, sensitivity and spatial resolution described in Section 3 will lead to new revelations. 

\begin{figure}
\includegraphics[scale=.25]{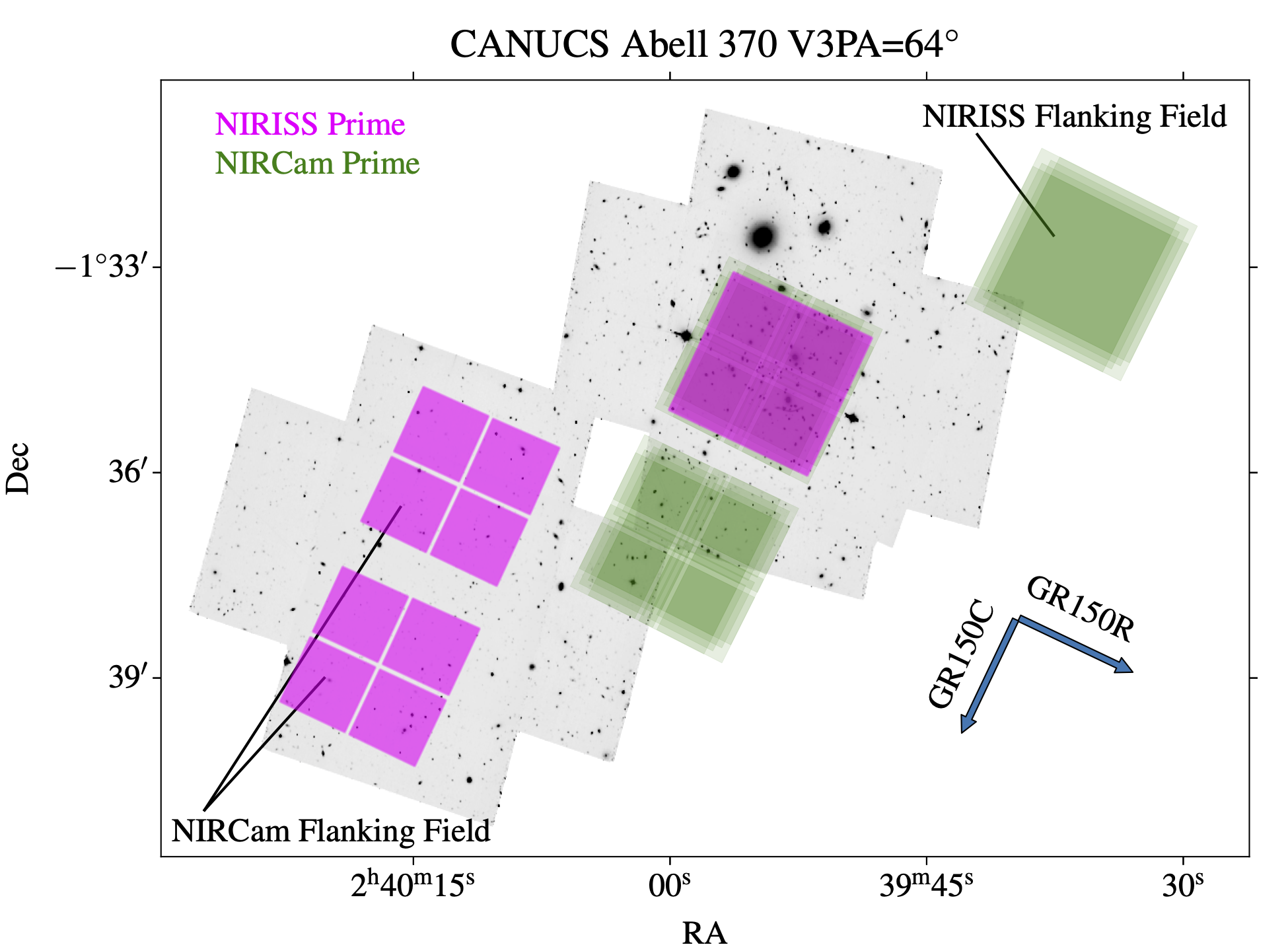}
\caption{Field layout for the CANUCS observations of Abell 370. The background greyscale is the BUFFALO ACS F814W image. NIRISS and NIRCam module B are centered on the cluster core. The NIRCam observations are made with a slightly larger dither pattern (INTRAMODULEX 6) and a small position offset so that NIRCam images the region outside the NIRISS imaging field where spectra are dispersed onto the detector. The NIRISS dispersion directions for the two gratings are shown with arrows. The NIRCam prime dither pattern is clearly visible (note that only the NIRCam Short Wave detectors, not Long Wave, are shown for clarity). When NIRISS is prime the dithers are smaller and barely perceptible on this plot. The full NIRCam Flanking Field is covered by BUFFALO optical imaging. The three NIRSpec MSA configurations in the field are not shown as their exact positions and orient angle will be defined later.
\label{fig:abell370}}
\end{figure}

The CAnadian NIRISS Unbiased Cluster Survey, or CANUCS\footnote{\url{http://canucs-jwst.com}}, will use about 200 hours of observatory time with the NIRISS, NIRCam and NIRSpec instruments. The NIRISS and NIRCam observations are performed in parallel at the same observatory orientation, so each target has a central field and two flanking fields, one observed with NIRISS and one with NIRCam, on either side. An example field layout for the CANUCS NIRISS and NIRCam prime and parallel observations is shown in Figure \ref{fig:abell370}. The central target fields for CANUCS (Table \ref{table:canucstargets}) are massive galaxy clusters chosen for their excellent gravitational lensing properties and extensive existing multi-wavelength data. Lensing cluster fields were chosen over non-cluster fields in order to take advantage of the lensing magnification, enabling the study of fainter and smaller galaxies. A lower limit of $z>0.35$ for target clusters was adopted to limit contamination from cluster galaxies and intracluster light, while also ensuring that most of the highly-magnified image plane is located within a single NIRISS field. Three of the clusters have extensive Hubble optical and NIR imaging from the Hubble Frontier Fields Project \citep{Lotz:2017} and the wider area BUFFALO survey \citep{Steinhardt:2020}. The other two clusters will be observed with 28 orbits each of ACS imaging in Hubble Cycle 29 program ID:16667 (PI: M. Brada\v{c}), expected to be completed before the CANUCS JWST observations of these fields.

Each cluster field in CANUCS will be observed with NIRISS with both orthogonal grisms (GR150C and GR150R) in the three filters; F115W, F150W and F200W. Total spectroscopic exposure times will be 9.7\,ks per configuration, resulting in sensitivities similar to those shown in Figs. \ref{fig:grismsnr} and \ref{fig:grismlinesnr}. Galaxy spectra uncontaminated in both grisms will have sensitivity of a factor of $\sqrt2$ better.

Grism spectroscopy in the NIRISS flanking fields will cover a larger volume than the cluster core fields (due to lower magnification) and therefore increase the statistics for more luminous galaxies. The NIRCam flanking fields will be imaged in 14 wide and medium filters: F090W, F115W, F140M, F150W, F162M, F182M, F210M, F250M, F277W, F300M, F335M, F360M, F410M, F444W (10 filters with 2.9 hours total exposure and 4 filters with 1.6 hours). With these data one can determine accurate photometric redshifts, strong emission line equivalent width distributions, and emission line, star formation rate and stellar mass maps \citep{Sorba:2018} up to redshift 9. 

The main science goal of CANUCS is to understand the evolution of low mass ($10^7\ltsimeq M_{*} \ltsimeq 10^9 M_{\sun}$) galaxies across cosmic time. Low mass galaxies typically reside in dark matter halos with mass $\ltsimeq 10^{11} M_{\sun}$ that have relatively low global star formation efficiency \citep{Moster:2013}. Possible reasons for the inability of gas to cool and form stars in these halos include feedback and outflows from supernovae, heating by the UV background and/or cosmic rays and heating and evaporation during reionization. By studying the physical properties of a large sample of such galaxies across cosmic epochs we aim to determine their evolutionary pathways.

\begin{figure*}
\includegraphics[scale=.55]{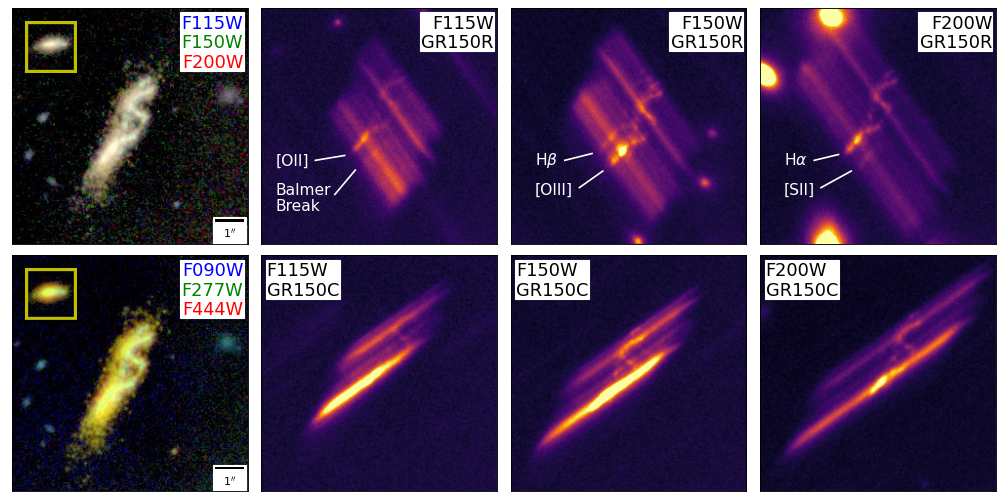}
\caption{Simulated CANUCS observations of a strongly lensed $z=2$ galaxy in the MACS~J0416.1-2403 field. All panels are $8\asec \times 8\asec$ with North up and East to the left. The two left panels show NIRISS (upper) and NIRCam (lower) 3-color images. The NIRISS filters include strong rest-frame optical emission lines, whereas the NIRCam filters shown are dominated by the stellar continuum. The inset image inside the yellow square is the unlensed galaxy. The six spectral panels show the NIRISS spectra of the two grisms combined with the three filters, all on the same intensity scale. For the orientation of this galaxy, the emission lines are more easily seen with the GR150R grism, however emission line maps can be generated for both grisms after continuum subtraction. The galaxy-shaped objects visible in some panels, particularly in GR150R with F150W and F200W, are zeroth order of other galaxies in the field.
\label{fig:lensedgal}}
\end{figure*}

In crowded fields such as strong-lensing clusters, grism spectroscopy is an efficient method to capture a spectrum of every galaxy in the field. By avoiding the biases of pre-selection of certain types of galaxies (as required for example with NIRSpec MOS) one obtains a sample of galaxies selected by their emission line properties. This is particularly important in the JWST era as previous work has shown that the equivalent width of emission lines in star-forming galaxies increases significantly both with increasing redshift and decreasing mass \citep{van-der-Wel:2011,Fumagalli:2012,Stark:2013,Smit:2014}. Extreme emission line galaxies (rest-frame equivalent width \gtsimeq 1000\AA) can be so dominated by lines that their continuum is undetectable. Such objects are easily revealed by grism spectroscopy (e.g. \citealt{Atek:2011}).

The small FWHM of NIRISS ($0.065 \asec$ at 2\,$\mu$m) enables high-resolution studies of distant galaxies, particularly when combined with the power of gravitational lensing. For a lensing magnification of a factor of five, this FWHM corresponds to a physical scale of 100\,pc for a $z=2$ galaxy. Therefore with CANUCS we will be able to generate resolved maps of physical properties (stellar mass, star formation rate, dust attenuation, nebular gas metallicity, etc.) for small, low-mass galaxies to understand their evolution including gas cycling with the circumgalactic medium. Individual star-forming regions the size of globular clusters will be measurable in some favorable lensing configurations (c.f. \citealt{Vanzella:2019}).

The GR150 grism blaze wavelength of 1.3\,$\mu$m was chosen to ensure the highest sensitivity for the Lyman-$\alpha$ transition redshifted to $z\approx 10$. This redshift is prior to the majority of cosmic hydrogen reionization, but late enough that the luminosity and space density of galaxies are both high enough to enable a comprehensive study with JWST. NIRISS grism spectroscopy is sensitive to faint Lyman-$\alpha$ emission lines, but also to the Lyman-$\alpha$ continuum   break in galaxies that do not show Lyman-$\alpha$ lines due to absorption in the IGM. With CANUCS we will search for `blind' Lyman-$\alpha$ emitters such as have been seen up to $z=6$ by MUSE \citep{Maseda:2018}, and determine the frequency of spatial offsets between line and continuum due to resonant scattering and obscuration (c.f. \citealt{Hoag:2019,Lemaux:2021}) that may complicate ground-based slit and NIRSpec micro-shutter-based Lyman-$\alpha$ spectroscopy. 

In Figure \ref{fig:lensedgal} we show a simulated CANUCS observation of a strongly-lensed galaxy at $z=2$. Mock JWST images of galaxies were created by combining the state-of-the-art, publicly available {\sc simba} cosmological simulation \citep{Dave:2019} with the open-source dust radiative transfer package {\sc powderday} \citep{Narayanan:2021}. We employed the $(25/h)^3$\,Mpc$^3$ {\sc simba} box with mass resolution of $\sim 10^6\,M_\odot$, meaning we can reasonably resolve galaxies with masses $>5\times 10^7\,M_\odot$. The minimum smoothing length for this simulation is $\sim 100$\,pc, sufficient for a strongly-lensed simulated galaxy. Spectra were assigned to each pixel of the full resolution simulation by matching the specific star formation rate to $z=2$ galaxies from the JAGUAR mock catalog \citep{Williams:2018}. 

The selected galaxy from the simulation has a star formation rate of $13\, M_\sun\,{\rm yr}^{-1}$ and lies on the average $z=2$ star-forming main sequence \citep{Santini:2017}. The simulated galaxy is inserted twice into the scene of the MACS\,J0416.1-2403 cluster, once at a position where it is subject to strong lensing (total magnification = 8.7) and once at the corner of the field with no lensing applied. Lensing is performed using the Lenstronomy package \citep{Birrer:2021} with the Zitrin NFW version 3 lensing model for this cluster \citep{Zitrin:2013}. The MACS\,J0416.1-2403 scene including these two galaxies was simulated using MIRAGE \citep{Hilbert:2019} with the CANUCS observational parameters for NIRCam imaging and NIRISS direct imaging and WFSS. The lensed galaxy has an observed F200W AB magnitude of 20.3. The raw data were passed through the CANUCS data processing pipeline to produce final data products. The dispersed spectra of the lensed galaxy in Figure \ref{fig:lensedgal} illustrate how well NIRISS users will be able to make emission line ratio maps of such galaxies (after continuum subtraction) for resolved studies of the ionization and metallicity of nebular gas.

\subsection{The NIRISS Survey for Young Brown Dwarfs and Rogue Planets}

Brown dwarfs are a natural byproduct of star formation. About a quarter of the objects found so far in nearby star forming regions are below the substellar threshold with estimated masses $<0.08\,M_{\odot}$ (e.g., \citealt{Andersen:2008,Muzic:2019}). Some of them should have planetary masses below the Deuterium burning limit at $M<0.013\,M_{\odot}$, corresponding to about 13$\,M_\mathrm{Jup}$ (\citealt{Pena-Ramirez:2012}, see review by \citealt{Luhman:2012}). Some of these objects will have been formed like stars, from the collapse of cloud cores, but others could be ejected (or \textit{rogue}) planets. Based on deep imaging of star forming regions, followed by spectroscopy of candidates, the number of free-floating objects with planetary masses has been estimated as 2-5 per 100 stars \citep{Scholz:2012a}. In young, dispersed associations, the number of such objects could be significantly higher \citep{Gagne:2017}. So far, however, none of the direct detection surveys are sensitive below a limit of $\sim 5\,M_\mathrm{Jup}$.

\begin{figure}
\includegraphics[scale=.61]{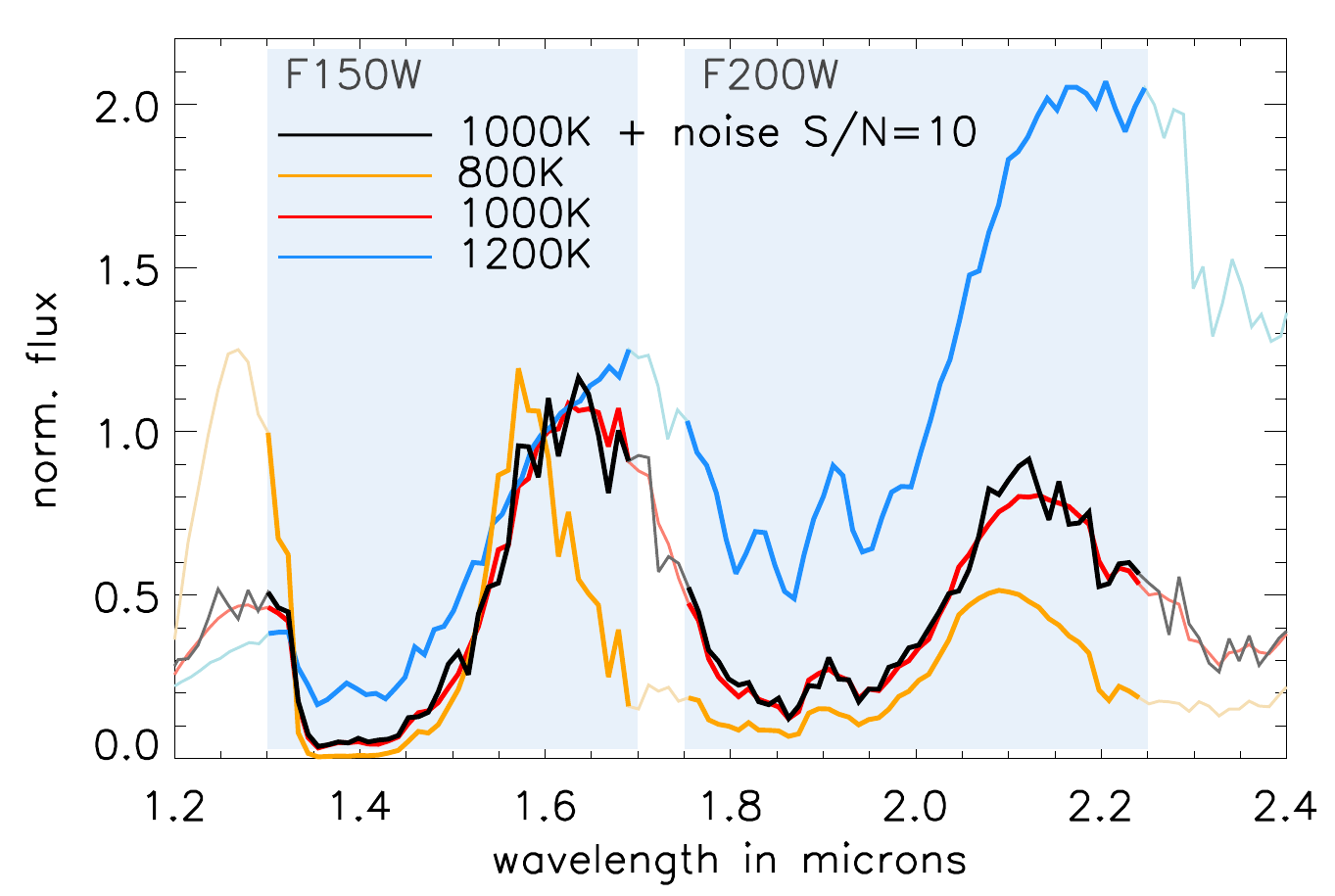}
\caption{Simulated spectral data for young ultracool objects. Model spectra from the BT-Settl series were smoothed to the WFSS resolution of $R=150$. For the input spectrum with a temperature of 1000\,K noise has been added to achieve the expected minimum S/N of 10.
\label{fig:bdspec}}
\end{figure}

\begin{figure*}
\includegraphics[scale=.655]{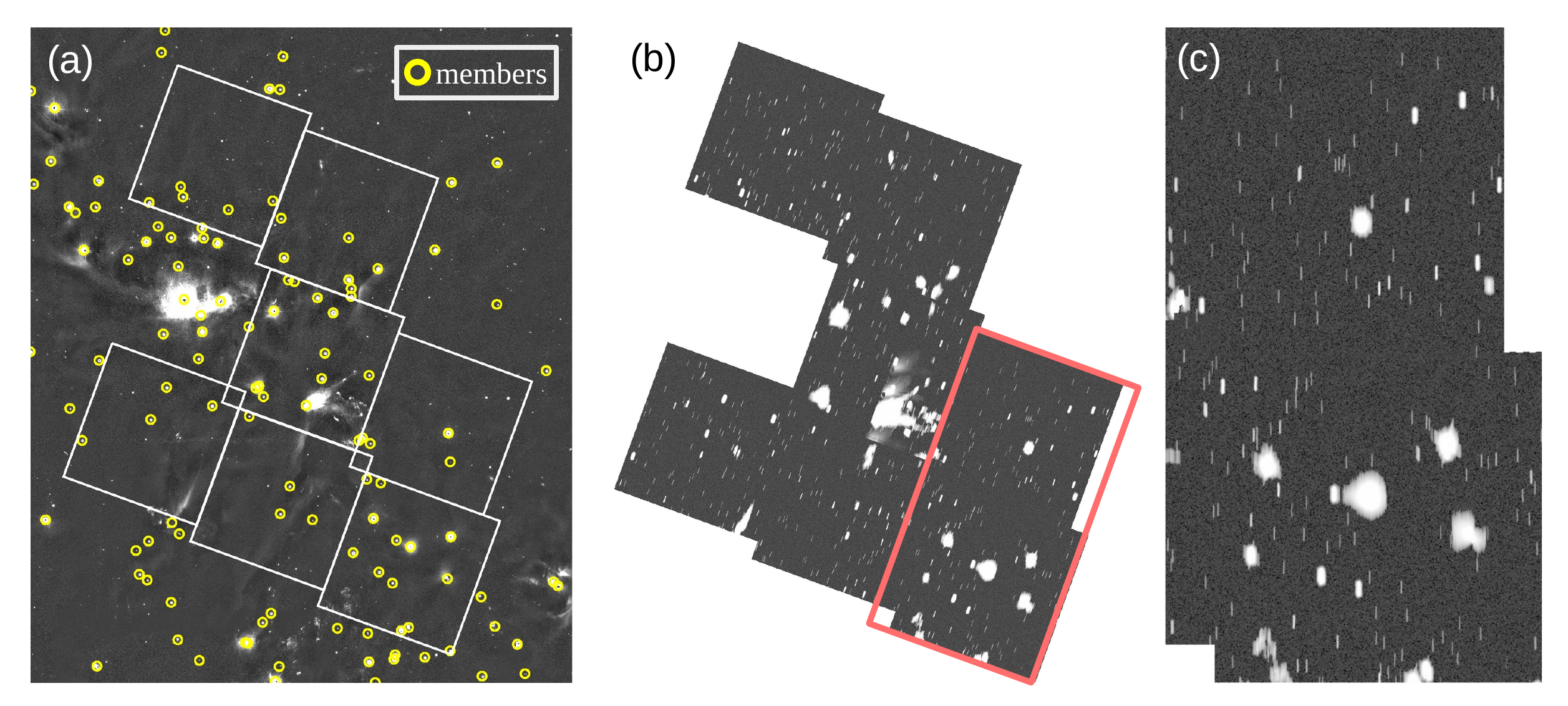}
\caption{(a) Subaru $K_s$-band image of the central part of NGC 1333, with the NIRISS WFSS mosaic overlaid. The yellow circles mark the spectroscopically confirmed cluster members from \cite{Luhman:2016}. (b) The \texttt{grizli} simulation with grism GR150C and filter F150W based on the original image to which faint sources expected to be detected by JWST were added. (c) Zoom into the two simulated WFSS fields marked by the red rectangle in panel (b).      
\label{fig:bdimage}}
\end{figure*}

In extinction-free nearby regions, brown dwarfs and free-floating planets separate sufficiently from background stars in broadband colors (typically red optical/NIR bands) and in proper motion to reliably select them and derive mass functions from imaging alone (e.g. \citealt{Pinfield:2000,Bejar:1999,Lodieu:2007}). In younger regions with significant reddening, however, young substellar objects mix in color-magnitude space with background stars and embedded young stars. Therefore, extensive spectroscopic follow-up is required \citep{Luhman:2003,Weights:2009,Wilking:2004}. 

Young brown dwarfs and giant planets have a characteristic NIR spectrum with broad, gravity-dependent absorption bands due to water and methane, which are easily distinguished from hotter photospheres with low-resolution ($R<1000$) and moderate signal-to-noise ratio (S/N$\approx$10).
The WFSS mode of NIRISS is an attractive and very efficient option for discovering an unbiased sample of these objects. WFSS has the obvious advantage of providing spectral information with sufficient resolution for the entire population in the targeted field, short-cutting the typical two-step strategy of photometry first, followed by spectroscopy for color-selected candidates. A spectroscopic survey with WFSS should be much less affected by incompleteness biases than color-selected samples. 

The NIRISS Survey for Young Brown Dwarfs and Rogue Planets\footnote{\url{https://www.stsci.edu/jwst/observing-programs/program-information?id=1202}} targets NGC 1333, a very compact young cluster. The cluster has a population of $>100$ known young stars within a diameter of 12 arcmin \citep{Scholz:2012,Luhman:2016}. The intra-cluster extinction is moderate, with a typical $A_v$ of 5 magnitudes towards the known stars and brown dwarfs. Contamination by scattered light from bright stars is not a severe issue. 

A Jupiter-mass object at age 1\,Myr has an effective temperature of $\approx 1000$\,K, using the AMES-Cond models \citep{Baraffe:2003}. At the distance of NGC 1333 of $\approx 300$\,pc such an object should have a brightness of $H\sim 22$ and $K\sim 21$\,mag\footnote{All magnitudes quoted in this subsection are in the Vega system.}. 
Using the JWST ETC (version 1.6), the total on-source exposure times of $\sim 2900$ and 3100\,s in the F150W (approximately $H$ band) and F200W (approximately $K$ band), respectively, with the GR150C grism, will achieve S/N=10 per spectral pixel. These two filters cover the most pronounced broadband absorption features of ultracool dwarfs. In Figure \ref{fig:bdspec} we show a BT-Settl model spectrum for a young 1000\,K source \citep{Baraffe:2015}, plus the expected noise, smoothed to the WFSS resolution $R=150$. For comparison, model spectra for a slightly warmer and a slightly cooler object are plotted, demonstrating the remarkable change in spectral shape as a function of temperature. 

The survey will cover a $3 \times 3$ pointing NIRISS mosaic of NGC 1333 (with 2 tiles removed, one affected by bright stars, the other with very few known cluster members), a total of about 34 square arcmin. These observations were simulated using \texttt{grizli}, with the MOIRCS/Subaru $K_s$-band image of the cluster \citep{Scholz:2009} as an input (Figure \ref{fig:bdimage}, left panel). This image reaches $K_s\sim 21\,$, and is complete down to $K_s\sim 19$. JWST will detect many more fainter objects, both in the cluster and in the field, and this was accounted for in the simulations, to verify that the resulting image would not be over-crowded.  From the histogram of brightness of the sources detected in Subaru images, the number of fainter sources expected to appear in the WFSS observations was estimated by extrapolation, and artificially added to the original image, evenly distributed across the field, assuming for simplicity a flat spectral template. In total, the simulated frames contain 350 real and 500 artificial sources. The simulations (Figure \ref{fig:bdimage}, middle and right panels) provide reassurance that crowding and overlapping spectra would not be a significant issue for this particular setup.

The spectra will distinguish between reddened background stars or embedded young stars and ultracool brown dwarfs - while the former have NIR spectra that are consistent with reddened blackbodies, the latter will feature strong absorption bands, as shown in Figure \ref{fig:bdspec}. Fitting the spectra with templates of young planetary mass objects or model spectra will determine the effective temperature and thus, in comparison with model isochrones, the mass. This project will determine the frequency of young free-floating objects with masses below the Deuterium burning limit down to a mass comparable to that of Jupiter.  


Star formation is expected to produce objects down to the opacity limit of fragmentation. Simulations suggest that this limit is in the range of a few Jupiter masses \citep{Bate:2012}, but it has not been found yet empirically. Dynamical simulations of planet ejections \citep{Parker:2012,van-Elteren:2019} suggest that significant numbers of ejected giant free-floating planets should exist (10 to 20 per 100 stars), but these model estimates are strongly dependent on assumptions about the architecture of the planetary systems and the clustered environment. The WFSS observations in NGC1333 can be used to compare with the existing simulations of star and planet formation, to establish the low-mass limit for objects to form like stars, and to account for the impact of ejections on the evolution of planetary systems.

\section {Summary}\label{sec:summary}

The wide field slitless spectroscopy mode of the NIRISS instrument provides a unique low-resolution spectroscopic option for users of the {\it James Webb Space Telescope}. Key highlights include the ease of use with no target acquisition, high multiplex factor and short spectral traces to minimize overlap. NIRISS WFSS has the potential to address science ranging from the solar neighborhood to the early Universe.

The filters covering the wavelength range $0.8\,-\,2.3\,\mu$m sample many important rest-frame NIR, optical and UV emission lines for galaxies at a range of redshifts plus diagnostic molecular features in ultracool dwarf atmospheres. 

In this paper we have detailed the pre-launch expected performance of the observing mode and some of the most important operational and calibration considerations for users. It is hoped that this information will encourage prospective users to be less daunted by this observing mode and consider how the mode can advance their science goals.

\begin{acknowledgments}

Special thanks to all our colleagues who have worked so hard over the last 20 years to design, build and test NIRISS. Thanks to Guido Roberts-Borsani for assistance with the gravitational lensing simulation. We acknowledge support from the
Natural Sciences and Engineering Research Council of Canada (NSERC) through a variety of its funding programs. This work is in part supported by Canadian Space Agency grant 18JWSTGTO1. CP, SR and KV are supported by the Canadian Space Agency under a contract with NRC Herzberg Astronomy and Astrophysics. KM acknowledges funding by the Science and Technology Foundation of Portugal (FCT), grants No. IF/00194/2015, PTDC/FIS-AST/28731/2017, UIDB/00099/2020.

\end{acknowledgments}

\facility{JWST (NIRISS)}


\bibliography{willott.bib}{}
\bibliographystyle{aasjournal}



\end{document}